\documentclass[11pt,twocolumn,preprintnumbers,amsmath,amssymb,nofootinbib,superscriptaddress,aps,prd,numberedappendix]{openjournal}
\usepackage{natbib}
\usepackage{newtxtext,newtxmath}
\usepackage{graphicx}
\usepackage{dcolumn}
\usepackage{bm}
\usepackage[dvipsnames]{xcolor}
\usepackage{xspace}
\usepackage{ulem}
\usepackage{lipsum}
\usepackage[colorlinks,linkcolor=red,citecolor=blue,urlcolor=blue ]{hyperref}
\usepackage{graphicx} 
\usepackage{multirow}

\newcommand{\resamp}{A_{\rm res}}
\newcommand{\Om}{\Omega_M}

\begin{document}
\title{Accuracy requirements on intrinsic alignments for Stage-IV cosmic shear}

\author{Anya Paopiamsap$^{1,*}$}
\author{Natalia Porqueres$^{1,*}$}
\author{David Alonso$^{1}$}
\author{Joachim Harnois-Deraps$^2$}
\author{C. Danielle Leonard$^2$}
\email{$^*$anyabua@gmail.com}
\affiliation{$^1$Department of Physics, University of Oxford, Denys Wilkinson Building, Keble Road, Oxford OX1 3RH, United Kingdom}
\affiliation{$^2$School of Mathematics, Statistics and Physics, Newcastle University, Newcastle upon Tyne, NE1 7RU, UK}
\date{\today}

\begin{abstract}
  In the context of cosmological weak lensing studies, intrinsic alignments (IAs) are one of the most complicated astrophysical systematics to model, given the poor understanding of the physical processes that cause them. A number of modelling frameworks for IAs have been proposed in the literature, both purely phenomenological or grounded on a perturbative treatment of symmetry-based arguments. However, the accuracy with which any of these approaches is able to describe the impact of IAs on cosmic shear data, particularly on the comparatively small scales ($k\simeq 1\,{\rm Mpc}^{-1}$) to which this observable is sensitive, is not clear. Here we quantify the level of disagreement between the true underlying intrinsic alignments and the theoretical model used to describe them that can be allowed in the context of cosmic shear analyses with future Stage-IV surveys. We consider various models describing this ``IA residual'', covering both physics-based approaches, as well as completely agnostic prescriptions. The same qualitative results are recovered in all cases explored: for a Stage-IV cosmic shear survey, a mis-modelling of the IA contribution at the $\sim10\%$ level produces shifts of $\lesssim0.5\sigma$ on the final cosmological parameter constraints. Current and future IA models should therefore aim to achieve this level of accuracy, a prospect that is not unfeasible for models with sufficient flexibility.
\end{abstract}

\maketitle
\section{Introduction}
  Intrinsic alignments (IAs) are intrinsic inherited shape correlations between neighbouring galaxies. These intrinsic alignments are related to galaxy formation processes and gravitational interactions between galaxies and their environment \citep{Heavens2000, Croft2000}. Multiple surveys have detected IAs to high significance \citep{Brown2002, Mandelbaum2006, Hirata2007, Joachimi2011, Singh2015, Johnson2019, Fortuna2021}. These non-cosmological shape correlations affect the cosmic shear estimates, and ignoring this effect can lead to significant cosmological biases \citep{Joachimi2010, Kirk2012, Krause2016}. 

  There are several models to describe IAs, including the non-linear tidal alignment \citep[NLA, ][]{Bridle2007}, the tidal alignment and tidal torquing model \citep[TATT][]{Blazek2019}, hybrid perturbative approaches \citep{Maion2023}, and the halo model \citep{Schneider2010, Fortuna2021b}. However, there is still a large uncertainty regarding the strength of IAs in typical cosmic shear samples. DES-Y1 \citep{Troxel2018}, KiDS \citep{Asgari2021} and HSC \citep{Hikage2019, Hamana2020} used the NLA model and reported non-zero values for the NLA amplitude, while DES-Y3 \citep{Secco2022} used both NLA and TATT, finding an NLA IA amplitude consistent with zero. In fact, the NLA amplitude seems to depend strongly on both galaxy type as well as the analysis choices made in cosmic shear studies (e.g. scale cuts, or choice of summary statistic -- see Table 2 of \cite{Asgari2021} or \cite{Tsaprazi2022} for a field-level approach). \cite{Samuroff2019} detected intrinsic alignments for early-type galaxies but found an amplitude consistent with zero for late-type galaxies with both NLA and TATT models. Some studies \citep{Blazek2015, Fortuna2021b, Troxel2018} found that the NLA model is disfavoured over more complex models, while \cite{Secco2022} found a mild preference for NLA over TATT.
  
  Neither do hydrodynamical simulations all agree on the best description to model IAs \citep{Samuroff2021}. While the galaxies in the IllustrisTNG simulation strongly disfavour tidal torquing and did not detect intrinsic alignment for blue galaxies \citep{Zjupa2020}, the {\tt Horizon-AGN} simulation shows IAs for high-redshift blue galaxies \citep{Codis2015}. Finally \cite{Blazek2019} showed that if the true physics are described by TATT, but the inference assumes the NLA model, the results are biased by several $\sigma$ for an LSST-like survey. The ability of different models to reproduce realistic IA signals on different scales is not completely clear. While there is evidence that NLA may be a reasonably good description up to $k\sim1\,h{\rm Mpc}^{-1}$ \citep{Shi2021}, there is also evidence that TATT, as well as other hybrid parametrisations, are able to significantly outperform it on small scales when describing halo alignments \citep{Maion2023}.The landscape of IA models is complex, their ability to recover a realistic IA signal in different regimes is unclear, and the expected amplitude of IAs compared to the cosmological shear signal is uncertain.

  In this paper, we aim to answer the question: how accurately must IA models reproduce the underlying IA signal on different scales to avoid causing a significant bias on cosmological parameters in cosmic shear analyses? Importantly, we will not concern ourselves with determining what model (NLA, TATT, halo model etc.) is more correct, but rather, what the maximum modelling residual for IAs is that future cosmic shear surveys can afford. Addressing this question is important for future ``Stage-IV'' surveys, such as the Vera C. Rubin Observatory Legacy Survey of Space and Time \citep[LSST, ][]{LSST2019}, the Euclid mission \citep{Laureijs2011Euclid}, or the Roman Space Telescope \citep{Spergel2015Roman}, given their significantly higher sensitivity compared to current datasets. While any of the phenomenological approaches described above may be able to reproduce a realistic IA signal at the $O(10\%)$ level (e.g. the typical precision of halo-based models for clustering statistics), a $O(1\%)$ accuracy requirement on both linear and non-linear scales (comparable to the accuracy with which the matter power spectrum itself must be known) is a significantly more challenging prospect.

Following this logic, in our analysis, we introduce perturbations of a given relative amplitude with respect to a known IA model (NLA or TATT) in the intrinsic alignment component of weak lensing angular power spectra, for an LSST-like survey. We then measure the corresponding bias induced on cosmological parameters ($\Om$ and $S_8\equiv\sigma_8\sqrt{\Om/0.3}$) when the perturbed data is analysed assuming a pure NLA or TATT model, and characterise this bias as a function of the amplitude of the perturbation, as well as the prescription used to generate it.

This paper is structured as follows. We present the intrinsic alignment models used in  Section \ref{sec:theory}. Section \ref{sec:meth} describes the methodology used to generate simulated data vectors with perturbed IA contamination, and to extract requirements for IA modelling accuracy from these data. We discuss the results in Section \ref{sec:results} and conclude in \ref{sec: conclusions}.

\section{Background theory} \label{sec:theory}
  In this section, we describe the IA contribution to cosmic shear power spectra and the different IA models considered in this work.

  \subsection{Intrinsic alignments contribution to power spectra}\label{ssec:theory.IA}
    Intrinsic alignments contribute to the shear power spectrum $C_{\gamma \gamma}(\ell)$ with two additive terms to the cosmological pure lensing signal (GG): the II term to model the correlation in the intrinsic shapes of physically close galaxies subject to the same tidal forces \citep{Catelan2001, Mackey2002} and the GI term to describe the correlation in shape between foreground galaxies and background sources due to the foreground mass \citep{Hirata2004}. The observable estimate of the shear angular power spectrum between redshift bins $i$ and $j$ is the sum of all three contributions,
    \begin{equation}
        C^{ij}_{\gamma \gamma} (\ell) = C^{ij}_{\mathrm{GG}}(\ell) + C^{ij}_{\mathrm{II}}(\ell)  + C^{ij}_{\mathrm{GI}}(\ell) +  C^{ij}_{\mathrm{IG}}(\ell).
    \end{equation}

    In the Limber approximation and assuming a flat Universe, the GG term is expressed in terms of the matter power spectrum as 
    \begin{equation}
        C^{ij}_\mathrm{GG} (\ell) = \int_0^{\chi_\mathrm{lim}} \frac{g^{i}(\chi)g^{j}(\chi)}{\chi^2}P_{\delta}(k_\ell, z)d\chi,
    \end{equation}
    where $k_\ell\equiv(\ell+1/2)/\chi$, $\chi_\mathrm{lim}$ is the horizon comoving distance, and the lensing kernel at each bin is given by
    \begin{equation}
        g^i(\chi ) = \frac{3}{2} \frac{H_0^2 \Om}{c^2} \frac{\chi}{a(\chi)} \int_\chi^{\chi_\mathrm{lim}} n^i(\chi') \frac{\chi'-\chi}{\chi'} d\chi'.
    \end{equation}
    The redshift distributions $n(z)$ are assumed to be defined such that $n(z)dz = n(\chi)d\chi$ and be normalised over the depth of the survey. 
    
    The Limber approximation of the II and GI terms are expressed in terms of the intrinsic alignment power spectra as
    \begin{equation}
        C^{ij}_\mathrm{II} (\ell) = \int \frac{n^{i}(\chi) n^{j}(\chi)}{\chi^2} P_\mathrm{II}(k, z) d\chi,
    \end{equation}
    \begin{equation}
        C^{ij}_\mathrm{GI} (\ell) = \int \frac{g^{i}(\chi) n^{j}(\chi)}{\chi^2}P_{\rm GI}(k, z)d\chi,
    \end{equation}
    where the power spectra $P_\mathrm{GI}$ and $P_\mathrm{II}$ are given by one of the intrinsic alignment models described below.

    \subsubsection{Non-linear alignment model}\label{sssec.theory.IA.nla}
      The non-linear alignment model \citep[NLA,][]{Bridle2007} assumes that the intrinsic galaxy shapes are linearly related to the local tidal field with a redshift-dependent scaling:
      \begin{align}
        P_{\rm GI}(k, z) &= A(z)P_{\delta}(k, z) \\
        P_{\rm II}(k, z) &= A^{2}(z)P_{\delta}(k,z) 
      \end{align}
      where the $P_{\delta}(k,z)$ here is the non-linear matter power spectrum and the normalisation is 
      \begin{equation}\label{eq: meth.nla.aia}
        A(z) = -A_1\bar{C}_{1}\rho_{{\rm crit}}\frac{\Om}{D(z)}\left(\frac{1 + z}{1 + z_{0}}\right)^{\eta_1}
      \end{equation}
      such that $A_1$ is an unknown scaling parameter that controls the strength of the intrinsic alignments and is left as a free parameter. $\eta_1$ controls the redshift evolution of the IA amplitude, and will also be left as a free parameter. $D(z)$ is the linear growth factor, and the normalisation constant is fixed to $\bar{C}_1 = 5 \times 10^{-14} M_{\odot}^{-1}h^{-2}{\rm Mpc}^3$. In this work, we fix the pivot redshift to $z_0=0.62$ \citep{DES:2018gxz}.

    \subsubsection{Tidal alignment and Tidal Torquing Model}\label{sssec.theory.tatt}
      The TATT model \citep{Blazek2019} is based on perturbation theory and combines tidal alignments (linear in the tidal field) and tidal torquing (quadratic in the tidal field). In the TATT model, the intrinsic galaxy shape $\gamma_{ij}^{I}$ is an expansion in the tidal field $s_{ij}$ and the density field $\delta$: 
      \begin{equation}
        \gamma^{I}_{ij} = C_1 s_{ij} + C_2\left( s_{ik}s_{kj} - \delta^{\rm K}_{ij}\frac{1}{3}s^2\right) + C_{1\delta}(\delta \times s_{ij}) + \dots,
      \end{equation}
      where $\delta^{\rm K}_{ij}$ is a Kronecker delta. $C_1$ describes the tidal alignment contribution and corresponds to the NLA model. $C_2$ captures the contribution from tidal torquing, and  $C_{1\delta}$ is a density-weighted contribution to the tidal alignment and, in the simplest case, is determined by the bias of the source sample $C_{1\delta} =  b_g C_1$. The components of the tidal field in Fourier space are given by
      \begin{equation}
        \Tilde{s}_{ij}(\boldsymbol{k}) = \left(\frac{k_j k_j}{k^2}  - \frac{1}{3} \delta^{\rm D}_{ij} \right) \Tilde{\delta}(\boldsymbol{k}) \label{eq:tidal_field}.
      \end{equation}

      In this work, we follow the parametrisation:
      \begin{align}
        C_1(z) &= - A_1\bar{C}_1 \rho_{\rm crit}\frac{\Om}{D(z)}\left( \frac{1 + z}{ 1 + z_0}\right)^{\eta_1}, \\
        C_2(z) &= 5A_2\bar{C}_1\rho_{\rm crit}\frac{\Om}{D^2(z)}\left(\frac{1+z}{1+z_0}\right)^{\eta_2},
      \end{align}
      where we consider $A_1$, $A_2$ and $A_{1\delta}\equiv b_gA_1$ free parameters to be inferred. For simplicity, we fixed $\eta_1=\eta_2$, assuming that both amplitudes have the same redshift evolution.

    \subsubsection{Extended NLA}\label{sssec.theory.IA.enla}
      While hydrodynamical N-body simulations find that $A_2$ is consistent with zero \citep{Samuroff2021}, the density-weighted term is physically motivated because we only observe IAs at the positions of galaxies, which are biased tracers of the dark matter distribution. This is modelled in the extended NLA model \citep[eNLA][]{Blazek2015, Blazek2019, HarnoisDeraps2022}, which reads simply
      \begin{equation}
        \gamma^{I}_{ij} = C_1 s_{ij} + C_{1\delta}(\delta \times s_{ij}).
      \end{equation}
      This is consistent with the TATT model for a torque term with null $A_2$ and reduces to the NLA by setting $b_g$ to zero. The density-weighted extension mostly affects the small scales \citep{Blazek2019} and has been detected in hydrodynamical simulations \citep{Hilbert2017, Samuroff2021}.
      \begin{figure*}
        \centering
        \includegraphics[width = 0.7\textwidth]{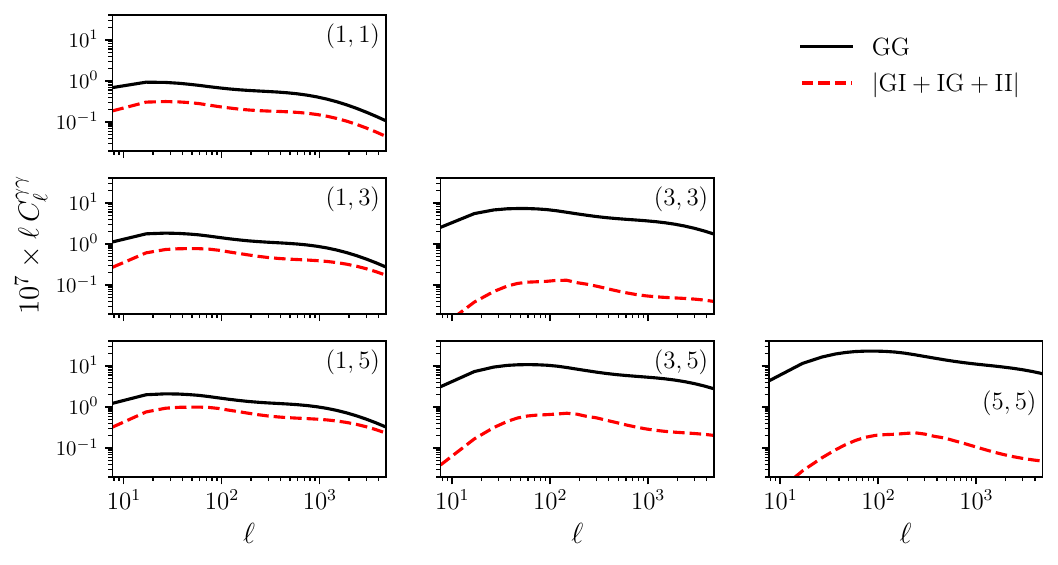}
        \caption{Shear power spectra involving the first, third, and fifth redshift bins shown in Fig. \ref{fig:tomobins} The black lines show the contribution from cosmic shear, while the red dashed lines show the contribution from all terms involving intrinsic alignments for an NLA model with $A_1=1$. The IA contributions are generally much smaller (between 1 and 2 orders of magnitude) than the cosmic shear term, except at the lowest redshifts, where the cosmic shear signal is significantly weaker. The constraining power on cosmological parameters is dominated by the higher redshift bins, where the signal is stronger and IAs weaker.}
        \label{fig:cls}
      \end{figure*}
    
    \subsection{The relative amplitude of intrinsic alignments}\label{ssec.theory.estimate}
      Having reviewed the theory behind some of the most common intrinsic alignment models, we can now try to produce a rough estimate of the main quantity we wish to calculate in this paper: the relative accuracy with which the underlying IA signal must be modelled to avoid significant biases in the final cosmological parameters.

      First, to get a heuristic idea of the relative IA contribution to the total cosmic shear signal, consider a Stage-IV weak lensing survey, such as the one described below in Section \ref{ssec:meth.simdata}. Figure \ref{fig:cls} shows the cosmic shear angular power spectra involving the first, third, and fifth redshift bins of this survey (see associated redshift distributions in Fig. \ref{fig:tomobins}). The black line shows the contribution from weak lensing, while the red dashed lines show the sum of all terms involving IAs for an NLA model with amplitude $A_1=1$. We can see that the IA contributions are generally subdominant to the lensing component by a factor $\sim10-100$, except at the lowest redshifts, where the lensing contribution is significantly weaker and both terms become comparable. Since the statistical power of the weak lensing signal is dominated by the higher redshift bins, where the lensing signal is higher, the contribution to the final cosmological constraints from the lowest redshifts is small in general. Therefore, as a conservative estimate, we expect the overall average IA contribution to be at least a factor $\sim10$ smaller than the weak lensing signal.

       Previous works have shown that, given the sensitivity of Stage-IV weak lensing surveys, the matter power spectrum sourcing the lensing signal, must be modelled with an accuracy of $\sim1\%$ out to scales $k\sim1\,{\rm Mpc}\,h^{-1}$ to avoid significant cosmological parameter biases \citep{Huterer2005,Hearin2012,Martinelli21}. Given this, if the total cosmic shear signal must be known to percent-level precision (a requirement driven by the statistical uncertainties of future surveys), and if IAs are on average a $\sim10\%$ contribution to that signal, we expect that a $\sim10\%$ accuracy in the modelling of the latter, on all scales used in the analysis, should suffice to ensure unbiased cosmological constraints.

       As discussed above, this is indeed a very rough estimate, since the relative amplitude of the IA signal depends strongly on both source redshift and on scale, and different scales and redshifts contribute differently to the final cosmological constraints. Our aim in this paper is therefore to refine this estimate with a more detailed analysis. Nevertheless, this back-of-the-envelope calculation gives us an intuition of what to expect in the below analysis.

\section{Analysis methods}\label{sec:meth}
    
  \subsection{Simulated survey} \label{ssec:meth.simdata}
    \begin{figure}
      \centering
      \includegraphics[width = \columnwidth ]{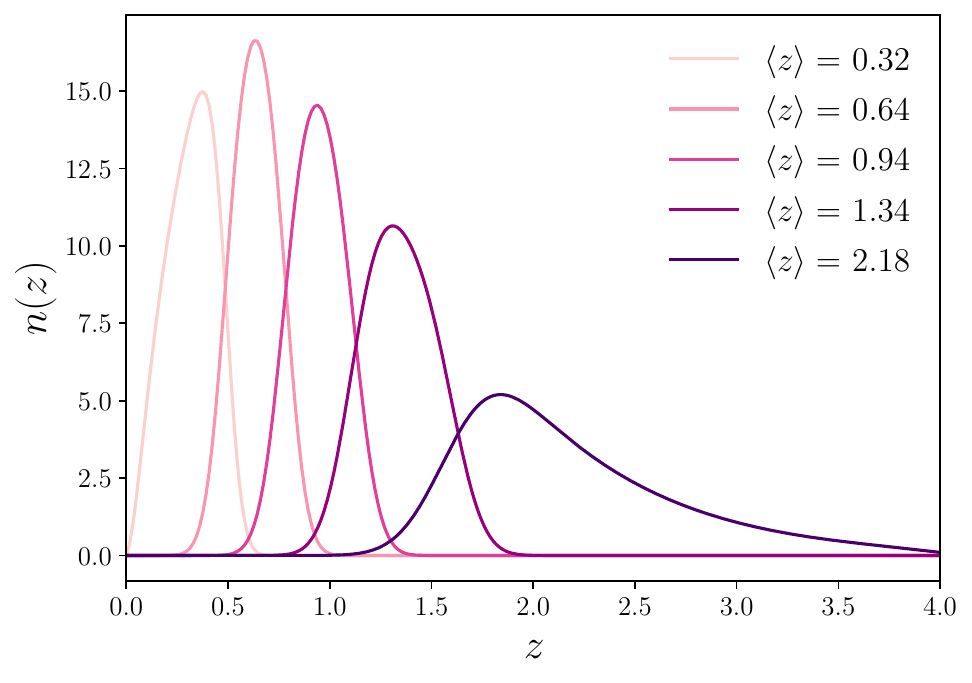}
      \caption{Redshift distribution of sources in each tomographic bin for the LSST-like survey described in Section \ref{ssec:meth.simdata}}
      \label{fig:tomobins}
    \end{figure}

    We generate simulated data vectors for an LSST-like survey. For this, we follow the description in the LSST Dark Energy Science Collaboration Science Requirements Document \citep{Mandelbaum2018LSST}. We divide the LSST 10-year shear sample, with a total number density of $26.5\,\,{\rm gal.}/{\rm arcmin}^2$, into 5 redshift bins, assuming Gaussian photo-$z$s with an uncertainty $\sigma_z=0.05(1+z)$. The bins were selected to have the same number of sources, and the resulting redshift distributions are shown in Fig. \ref{fig:tomobins}. The data vector consists of all auto- and cross-power spectra between the cosmic shear $E$-mode in these 5 redshift bins, including all angular multipoles in the range $\ell\in[30,2000]$, divided into 25 bandpowers as defined in \cite{Leonard2023}.
    
    The covariance matrix for this data vector was estimated including only the ``Gaussian'' or ``disconnected'' component, with the noise contribution to the auto-spectra calculated assuming an ellipticity scatter $\sigma_e=0.28$ \citep{Hilbert2017, Troxel2018}. This approach neglects the super-sample and other non-Gaussian contributions to the covariance matrix. This is a reasonable approximation given the large area covered by LSST \citep{Barreira2018}. In any case, neglecting these contributions would, if anything, under-estimate the final parameter uncertainties, thus leading to more stringent requirements on IA modelling accuracy than if they had been included. We assume a sky coverage of $f_{\rm sky}=0.4$.

    By default, all auto- and cross-spectra include an intrinsic alignment contribution assuming a baseline NLA model with IA amplitude $A_1=1$. This corresponds to the value preferred by the COSEBIs \citep{Schneider2010COSEBIs} analysis of the KiDS-1000 cosmic shear data \citep{Asgari2021}, which is the largest IA amplitude found by current cosmic shear surveys. This baseline IA contribution is then perturbed using the procedures outlined in Section \ref{ssec:meth.pert}. We will also study the impact on our conclusions of using a baseline model with a larger NLA amplitude, as well as using TATT instead of NLA as a baseline model.

    This sythetic dataset was generated assuming a $\Lambda$CDM cosmology with parameters $\Om=0.3$, $\Omega_b=0.05$, $h=0.7$, $n_s=0.96$, $\sigma_8=0.81$. The linear matter power spectrum was calculated using the {\tt CAMB} Boltzmann solver \citep{Lewis2000CAMB}, and then used to calculate the non-linear matter power spectrum using the HALOFIT prescription  \citep{Smith2003Halofit, Takahashi2012Halofit}. All theoretical calculations were carried out using the Core Cosmology Library \citep[CCL,][]{CCL}.

    We reiterate that all choices described above (purely Gaussian covariance, non-conservative scale cuts, a rather large baseline IA amplitude) were designed to enhance the relative impact of a mis-modelled IA signal, and thus lead to conservative accuracy requirement for IA models in next-generation surveys.

  \subsection{Perturbing IA models} \label{ssec:meth.pert}
      \begin{figure}
        \centering
        \includegraphics[width = \columnwidth]{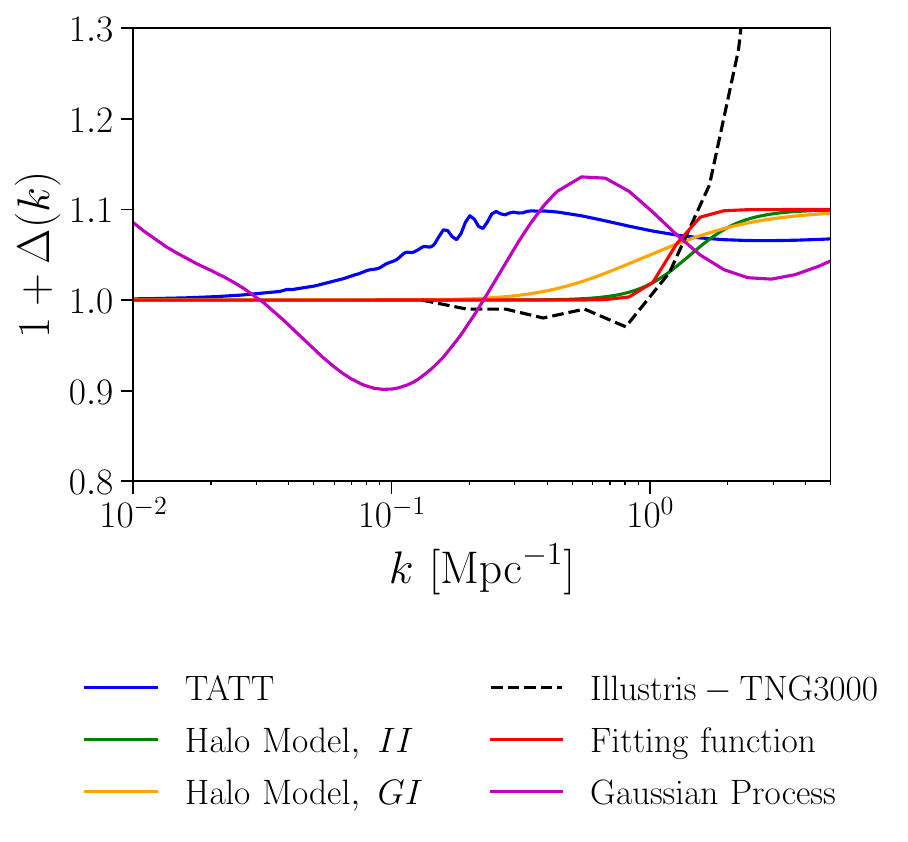}
        \caption{IA power spectrum residuals at $z=0$ generated by the different methods described in Section \ref{ssec:meth.pert} for a $\sim10\%$ residual amplitude.}
        \label{fig:a_ia_fit}
      \end{figure}
    \begin{figure*}
      \centering
      \includegraphics[width = 0.9\linewidth]{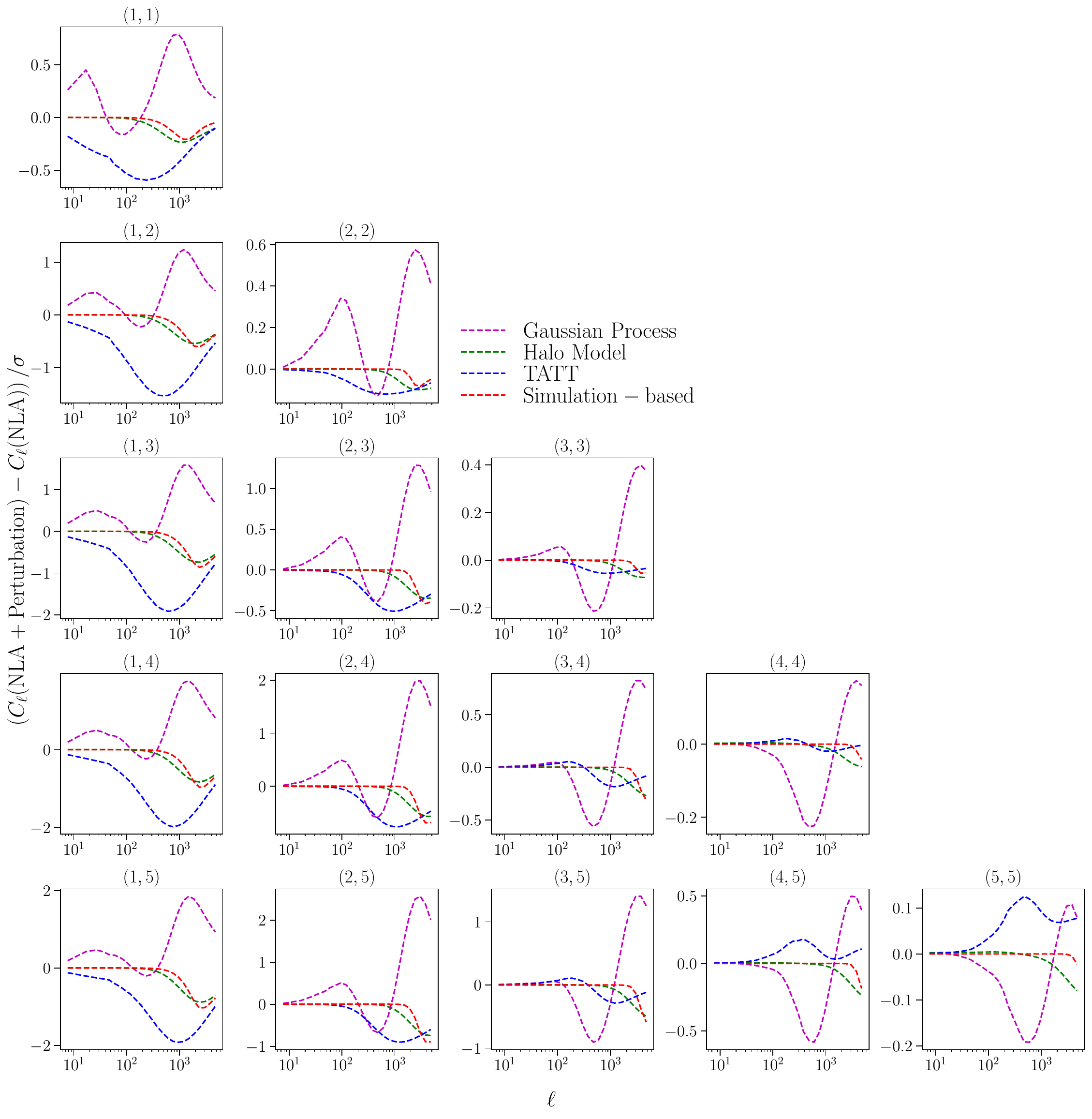}
      \caption{Deviation, relative to the statistial uncertainties, in the cosmic shear angular power spectra for the different types of IA residuals simulated here for a residual amplitude $\resamp=0.1$.}
      \label{fig:Delta_Cl_With_all_Perts}
    \end{figure*}
    As described in the previous section, to simulate the IA contribution in our synthetic data vector, we used NLA as a baseline model and added a perturbation to the IA power spectra to mimic potential residuals from the assumed IA model. We considered four types of residuals:
    \begin{enumerate}
        \item Gaussian-process residuals.
        \item Residuals based on simulations.
        \item TATT-like residuals.
        \item Halo model residuals.
    \end{enumerate}

    In all cases the methods used to generate the perturbations were designed to be able to control the relative amplitude of these perturbations, $\resamp$, allowing us to place a clear requirement on the relative amplitude of the IA modelling residuals. The synthetic data vectors thus perturbed were then analysed assuming (unperturbed) NLA, eNLA and TATT models to quantify their robustness to physical deviations from the model as a function of the amplitude of these deviations. We describe the methods used to add these types of perturbations in the next sections. Fig. \ref{fig:a_ia_fit} shows examples of these simulated residuals at the level of the three-dimensional IA power spectra for the various models used in this work, while Fig. \ref{fig:Delta_Cl_With_all_Perts} shows the impact of these deviations at the level of the angular power spectra as a fraction of the statistical uncertainties, in both cases assuming a 10\% relative residual amplitude.

    \subsubsection{Gaussian Process residuals} \label{sssec:meth.pert.gp}
      We first make use of Gaussian processes (GPs) to generate perturbations on the IA power spectra that modify both the scale dependence of the IA component and its redshift evolution, while ensuring a reasonable degree of smoothness in both properties.

      Concretely, we write the perturbed IA power spectra as:
      \begin{align}
        &P_{\rm GI}(k,z)=P_{\rm GI}^{\rm bl}(k,z)\left[1+\Delta_{\rm GI}(k,z)\right],\\
        &P_{\rm II}(k,z)=P_{\rm II}^{\rm bl}(k,z)\left[1+\Delta_{\rm II}(k,z)\right],\\
      \end{align}
      where the relative perturbations $\Delta_{\rm GI}(k,z)$ and $\Delta_{\rm II}(k,z)$ with respect to the baseline IA power spectra ($P_{\rm GI}^{\rm bl}(k,z)$ and $P_{\rm II}^{\rm bl}(k,z)$) were generated as realisations of a 2-dimensional Gaussian process with zero mean and a covariance matrix of the form:
      \begin{align}\nonumber
        K(k, k', a, a') &= \resamp^2\exp\left[{-\frac{1}{2}\left(\frac{{\rm log}(k/k')}{ \ell_{k}}\right)^2}\right] \\& \,\,\,\,\,\,\times\,\,\exp\left[{-\frac{1}{2}\left(\frac{a - a'}{ \ell_{a}}\right)^2}\right] + \sigma_{\rm noise}\delta_{ij}.
      \end{align}
      Here $\resamp^2$ is the signal variance, which controls the amplitude of the generated perturbations. $\ell_k$ and $\ell_a$ are the correlation lengths of the GP over log-wavenumber $\log k$ (with $k$ in units of ${\rm Mpc}^{-1}$), and scale factor $a\equiv 1/(1+z)$. $\sigma_{\rm noise}$ is a small noise jitter component that is included to regularise the covariance matrix and ensure it is invertible, but has otherwise no practical impact on the properties of the final GP realisations. The correlation lengths were fixed to $\ell_k= 1.0$ and $\ell_a= 0.1 $. We use the covariance amplitude $\resamp$ as the parameter that quantifies the relative amplitude of the IA residuals. Note, however, that a given GP realisation may at certain points deviate from zero by more than this amplitude (typically, around $\sim1/3$ of all GP points will deviate by more than 1$\sigma$).

      This approach generates perturbations on the IA power spectra that contribute to all scales and redshifts with the same variance (i.e. not favouring small scales over large scales). In this sense, this is arguably the most conservative of the perturbation models used (e.g. all other methods generate perturbations mostly on smaller scales, leaving large scales untouched). At the same time, GPs generate the least ``physical'' perturbations, since there is no obvious astrophysical process that could give rise to arbitrary deviations from the matter power spectrum on large scales. {\it A priori}, more physically-motivated perturbations could be mimicked by variations in cosmological parameters (e.g. changing the high-$k$ slope of the matter power spectrum, \cite{Giblin2019}), and thus could potentially lead to larger parameter biases for the same perturbation amplitude. This motivates the other approaches described below. 
    
    \subsubsection{Perturbation based on simulations} \label{sssec:meth.pert.sim}
      We generated a physically-motivated perturbation to the NLA IA power spectrum by mimicking the measurements of $P_{\rm GI}(k)/P_{\delta}(k)$ from the Illustris-TNG300 hydrodynamical simulation \citep{Shi2021}. Regarding the scale dependence of $P_{\rm GI}$, \cite{Shi2021} found that, on large scales, the NLA model is a reasonable description, but differences arise on small scales $k\sim1\,{\rm Mpc}^{-1}$. Since, within the scale cuts used here, our data vector is sensitive mostly to scales $k\lesssim 2\,{\rm Mpc}^{-1}$, we use the following sigmoid fitting function to characterise the scale dependence of $\Delta_{\rm GI}$ from the measurements of \citep{Shi2021}:
      \begin{equation}\label{eq:fitfunc}
        \Delta_\mathrm{GI}(k) = \frac{\resamp}{1+e^{(k_p-k)/\Delta k}},
      \end{equation}
      where $\resamp$ is the amplitude of the perturbation, $\Delta k$ describes the range of scales it affects, and $k_p$ denotes the pivot scale at which it becomes significant. In the absence of a better model, we generate the perturbation to the II power spectrum simply as:
      \begin{equation}\label{eq:simii}
        1+\Delta_\mathrm{II}(k) = \left[1 + \Delta_\mathrm{GI}(k)\right]^{2}.
      \end{equation}
      A comparison of Eq. \ref{eq:fitfunc} to the digitized data from \cite{Shi2021} is shown in Fig. \ref{fig:a_ia_fit} for the best fitting parameters $\resamp=0.1$, $\Delta k = 0.11\,\,{\rm Mpc}^{-1}$ and $k_p=1.2\,\,{\rm Mpc}^{-1}$. 
      
      It is worth noting that the value of $\Delta_{\rm GI}(k)$ measured by \cite{Shi2021} grows towards increasingly larger $k$. Since we wish to control the maximum amplitude of the IA perturbation to the baseline model, the fitting formula of Eq. \ref{eq:fitfunc} is able to recover the results of \cite{Shi2021} up to $k\sim1.5\,{\rm Mpc}^{-1}$ when setting $\resamp=0.1$ (i.e. 10\% deviation from NLA), and then smoothly asymptotes towards a constant value. The functional form of Eq. \ref{eq:fitfunc} should therefore not be taken to represent realistic perturbations on arbitrarily small scales, but only as a well-educated toy model for the scale dependence on scales close to those where the true IA power spectrum starts departing from the NLA prediction. Thus, although we will present fiducial results for the best fit values of $(k_p,\Delta k)$ quoted above ($1.2\,\,{\rm Mpc}^{-1}$ and $0.11\,\,{\rm Mpc}^{-1}$, respectively), we will also make use of Eq. \ref{eq:fitfunc} to explore the dependence of our results on the scale $k_p$ at which a deviation in the IA power spectrum starts, and the range of scales over which it is relevant.

    \subsubsection{Halo model} \label{sssec:meth.pert.hm}
      The halo model provides another principled approach to generate physical deviations with respect to a baseline IA model. In particular, we use the prescription of \cite{Fortuna2021}, which provides a halo model parametrisation for the satellite contribution to intrinsic alignments. This contribution affects only the 1-halo term of the full halo model description, with the 2-halo term, proportional to the matter power spectrum, being given exactly by the NLA model. The perturbation to the ${\rm GI}$ IA power spectrum could therefore be simply calculated as
      \begin{equation}
        \Delta_{\rm GI}(k,z)\propto \frac{P_{\rm GI}^{\rm sat}(k,z)}{P_{\rm m}(k,z)},
      \end{equation}
      where $P_{\rm GI}^{\rm sat}(k,z)$ is the 1-halo prediction of \cite{Fortuna2021}. As in the simulation-based model of the previous section, this prescription would give rise to a perturbation that grows arbitrarily large on sufficiently small scales. To keep the total amplitude of the IA perturbation under control, we therefore calculate it as:
      \begin{align}
        &\Delta_{\rm GI}(k,z)=\resamp\,{\rm Softmax}\left[\frac{P_{\rm m}(k_p,z)}{P_{\rm GI}^{\rm sat}(k_p,z)}\frac{P_{\rm GI}^{\rm sat}(k,z)}{P_{\rm m}(k,z)}\right],\\
        &\Delta_{\rm II}(k,z)=\resamp\,{\rm Softmax}\left[\frac{P_{\rm m}(k_p,z)}{P_{\rm II}^{\rm sat}(k_p,z)}\frac{P_{\rm II}^{\rm sat}(k,z)}{P_{\rm m}(k,z)}\right],
      \end{align}
      where ${\rm Softmax}(x)\equiv x/(1+|x|)$ ensures that $x$ never exceeds 1, and $\resamp$ is a free parameter that controls the relative amplitude of the IA residuals (e.g. $\resamp=0.1$ corresponds to $10\%$ relative deviations with respect to the baseline IA model).

      The halo model predictions for $P^{\rm sat}_{\rm GI}(k,z)$ and $P^{\rm sat}_{\rm II}(k,z)$ were generated using a satelite profile amplitude $a_{1h}=10^{-3}$, a power-law index $b=-2$ (see \cite{Fortuna2021} for a full description of these parameters), and the best-fit halo occupation distribution parameters of \cite{Nicola2020} for the HSC DR1 cosmic shear sample. Fig. \ref{fig:a_ia_fit} shows, in green and orange, the relative perturbations to the ${\rm II}$ and ${\rm GI}$ power spectra respectively.

    \subsubsection{TATT-like perturbations} \label{sssec:meth.pert.tatt}
      The TATT model itself can be used to generate perturbations with respect to a purely NLA baseline IA power spectrum. Although this approach would not allow us to explore the impact of IA power spectrum residuals when using a TATT parametrisation in the parameter inference stage of the analysis, it can be used to study the impact of IA residuals with a scale dependence that is significantly different to that induced through the simulation-based and halo model approaches, but which are, nevertheless, physically motivated. This is because, since TATT is a perturbative model, it causes deviations with respect to NLA on mildly non linear scales ($k\sim0.2\,{\rm Mpc}^{-1}$), and not only on non-linear ones ($k\sim 1\,{\rm Mpc}^{-1}$).

      To do this, we choose arbitrary values of the higher-order IA parameters $A_2 = 0.02$ and $A_{1\delta} = 0.02$ that cause a given maximum deviation with respect to our baseline IA power spectra with $A_1=1$ within the range of scales $k\lesssim 2\,{\rm Mpc}^{-1}$ that our analysis is sensitive to. The values of $\Delta_{\rm GI}(k,z)$ and $\Delta_{\rm II}(k,z)$ thus calculated are then modified by a free multiplicative factor $\resamp$ that controls the maximum relative deviation allowed. Fig. \ref{fig:a_ia_fit} shows, in blue, the TATT perturbation to the ${\rm II}$ power spectrum for a maximum deviation of $\resamp=0.1$ (i.e. $10\%$ relative perturbation).

  \subsection{Quantifying IA modelling requirements} \label{ssec:meth.like}
    \begin{table}
    \centering
    \begin{tabular}{|l l l|} 
    \hline
     & Parameters               & Prior          \\ 
    \toprule
    \multirow{2}{*}{{\bf Cosmology}}           & 
    $\Om$               & $U(0.05,0.7)$  \\ 
    & $\sigma_8$               & $U(0.5,1.2)$   \\ 
    \hline
    \multirow{2}{*}{\bf NLA}           & 
    $A_1$                    & $U(-5,5)$      \\ 
    & $\eta_1$              & $U(-5,5)$      \\ 
    \hline
    \multirow{4}{*}{\bf TATT}           

    & $A_1$                    & $U(-5,5)$      \\ 
    & $A_2$                    & $U(-5,5)$      \\ 
    & $A_d$                    & $U(-5,5)$      \\ 
    & $\eta_1 = \eta_2 = \eta$ & $U(-5,5)$      \\
    \hline
    \end{tabular}
    \caption{Table summarising the priors used in the likelihood for the cosmology, NLA and TATT parameters.}
    \label{tab:priors}
    \end{table}

    To quantify the accuracy requirements of intrinsic alignment models, we analysed the 4 datasets described above, assuming an unperturbed NLA, eNLA or TATT model. We used a Monte Carlo Markov Chain (MCMC) framework to analyse the perturbed synthetic data and infer $\Om$, $\sigma_8$ and the IA parameters: $\eta_1, A_1$ for NLA and $\eta_1, A_1, A_2, A_d$ for TATT. Although this is a limited cosmological parameter space, $\Om$ and $\sigma_8$ are the $\Lambda$CDM parameters to which cosmic shear is by far the most sensitive. We therefore expect this admittedly simplistic analysis scenario to provide a reasonable testing ground for the question at hand. We used the MCMC framework of {\tt cobaya} \citep{Torrado2021} and assume a Gaussian likelihood of the form
    \begin{equation}
      -2\log p({\bf d}|{\bf q})=\chi^2\equiv({\bf d}-{\bf t}({\bf q}))^T\,{\sf C}^{-1}({\bf d}-{\bf t}({\bf q}))+K,
    \end{equation}
    where ${\bf d}$ is the data vector (i.e. the full set of shear auto- and cross-spectra), ${\bf t}({\bf q})$ is the theoretical prediction within the model used, depending on parameters ${\bf q}$, ${\sf C}$ is the covariance matrix, and $K$ is an arbitrary constant. As described in Section \ref{ssec:meth.simdata}, ${\sf C}$ is computed considering only the ``Gaussian'' or ``disconnected'' contribution. We impose flat priors on all parameters, as described in Table \ref{tab:priors}.

    We will use two different metrics to evaluate the impact of IA modelling residuals:
    \begin{itemize}
      \item {\bf Relative bias.} For a given parameter $Q$, we extract the mean $\bar{Q}$ and standard deviation $\sigma_Q$ from the MCMC chains, and calculate the bias as a fraction of the statistical uncertainties:
      \begin{equation}
        \frac{\Delta Q}{\sigma_Q}\equiv\frac{|Q-Q_{\rm true}|}{\sigma_Q},
      \end{equation}
      where $Q_{\rm true}$ is the value used to generate the data.
      \item {\bf Goodness of fit.} We record the minimum $\chi^2$ value, and use it to quantify the ability of the IA model assumed to describe the perturbed data\footnote{Or, in other words, to determine whether the IA residual would be detectable by virtue of the model failing to provide a good description of the data.}. Note that the data vector used in our likelihood is noiseless. Thus, the minimum $\chi^2$ found must be interpreted as the excess $\chi^2$ over the mean $\chi^2$ expected for a multivariate random variable with $N_{\rm dof}=N_d-N_p$ degrees of freedom (with $N_d=375$ the number of elements in our data vector, and $N_p$ the number of free parameters) Thus, in what follows, when quoting the probability-to-exceed (PTE) of a given $\chi^2$, we refer to the PTE associated with a total chi-squared of $\chi^2+N_{\rm dof}$. This was shown to be a good approximation by \cite{Nicola2023}.
    \end{itemize}
        
\section{Results} \label{sec:results}
  \subsection{Gaussian-process perturbations}\label{ssec:res.gp}
    Using the Gaussian process model for the IA residuals described in Section \ref{sssec:meth.pert.gp}, we generated sets of data vectors with perturbation amplitudes of $\resamp=1\%,\,2\%,\,5\%,$ and $10\%$ with respect to the true IA power spectrum. For each perturbation amplitude, we generated 10 GP realisations of the perturbed data vector. We then analysed each of them independently, resulting in 10 MCMCs, and computed the average bias of the parameters as
    \begin{equation}\label{eq: paramovererror}
        \frac{\Delta Q}{\sigma_{Q}} \equiv \frac{\langle |Q - Q_{\rm true}|\rangle}{\sigma_{Q}}, 
    \end{equation}
    where the $\langle \cdot \rangle$ denotes an ensemble average across the 10 results.
    
    \begin{figure}
        \centering
        \includegraphics[width = \columnwidth]{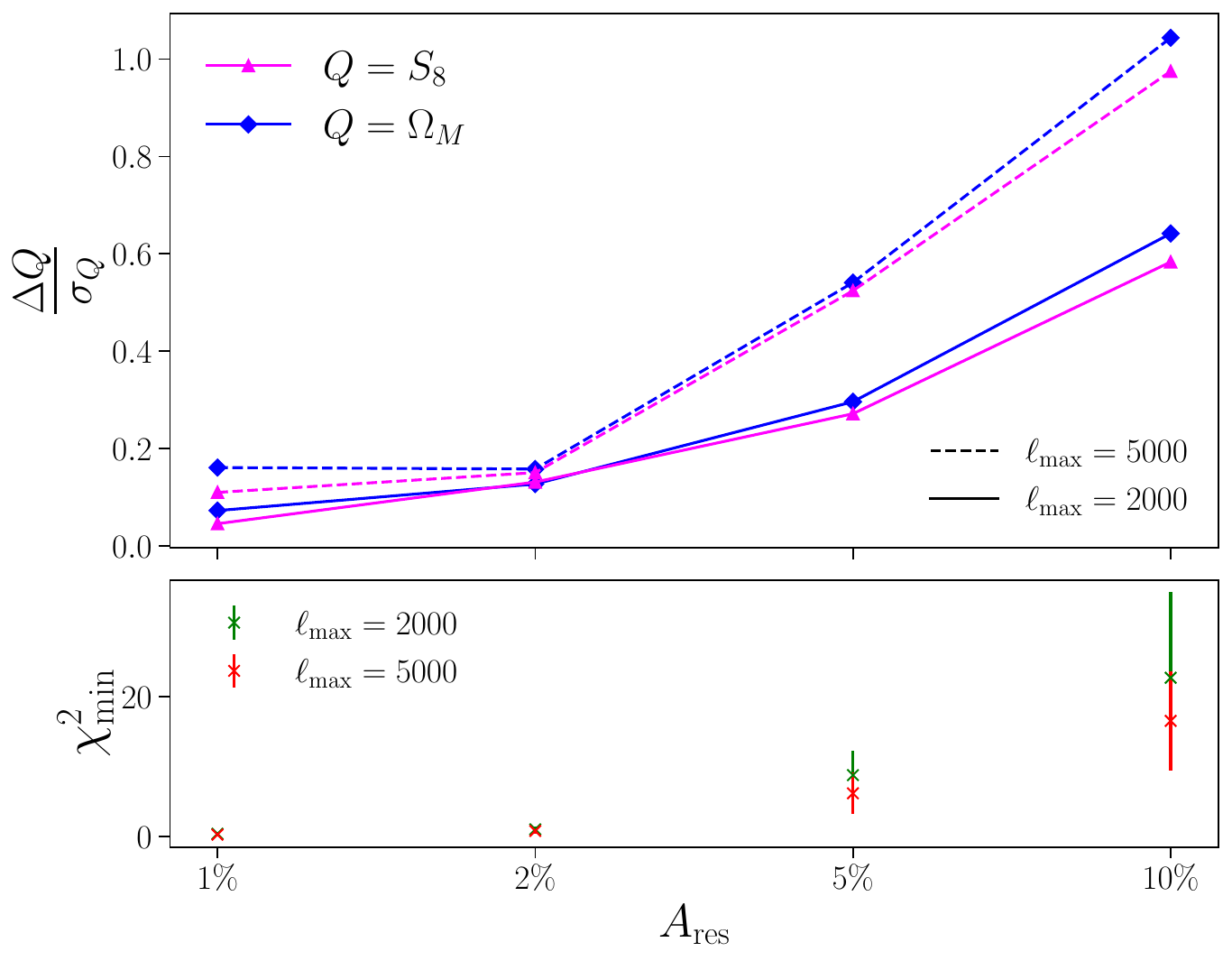}
        \caption{Average bias of $\Om$ and $S_8$ from NLA data with Gaussian-process perturbations when assuming an (unperturbed) NLA model in the analysis, as a function of the residual amplitude $\resamp$. The bottom panel shows the minimum $\chi^2$ achieved in each case, as a goodness-of-fit metric.}
        \label{fig:gaussian-NLA}
    \end{figure}

    Figure \ref{fig:gaussian-NLA} shows the average relative bias of $\Om$ and $S_8$ when analysing these data sets assuming an NLA model. While the bias increases with the amplitude of the perturbation, the bias stays below $\sim0.6\sigma$ for residuals with an amplitude  $\resamp<10\%$. Additionally, we explore the scale dependence of the average relative bias. As indicated by the dashed lines, we increased the maximum angular multipole $\ell_{\rm max}$ to 5000. We find that the average relative biases on these parameters are affected by the choice of scale cuts, with the maximum bias increasing from  $\sim0.6\sigma$ to $\sim1\sigma$ at $\resamp =10\%$, after doubling $\ell_{\rm max}$. Thus, scale cuts up to $\ell_{\rm max} = 2000$ can be considered a conservative but sensible choice for the analysis, given the balance between the validity of the NLA model and the overall IA contribution on small scales. It is worth noting that other astrophysical systematics, such as baryonic effects, also become relevant on small scales, and should be taken into account and marginalised over. This would likely lead to an increase in the statistical uncertainties of cosmological parameters, thus reducing the relative impact of IA modeling residuals from that reported here (the exact impact of the interplay between both astrophysical systematics should be studied in detail, however).

    \begin{figure}
        \centering
        \includegraphics[width=\columnwidth]{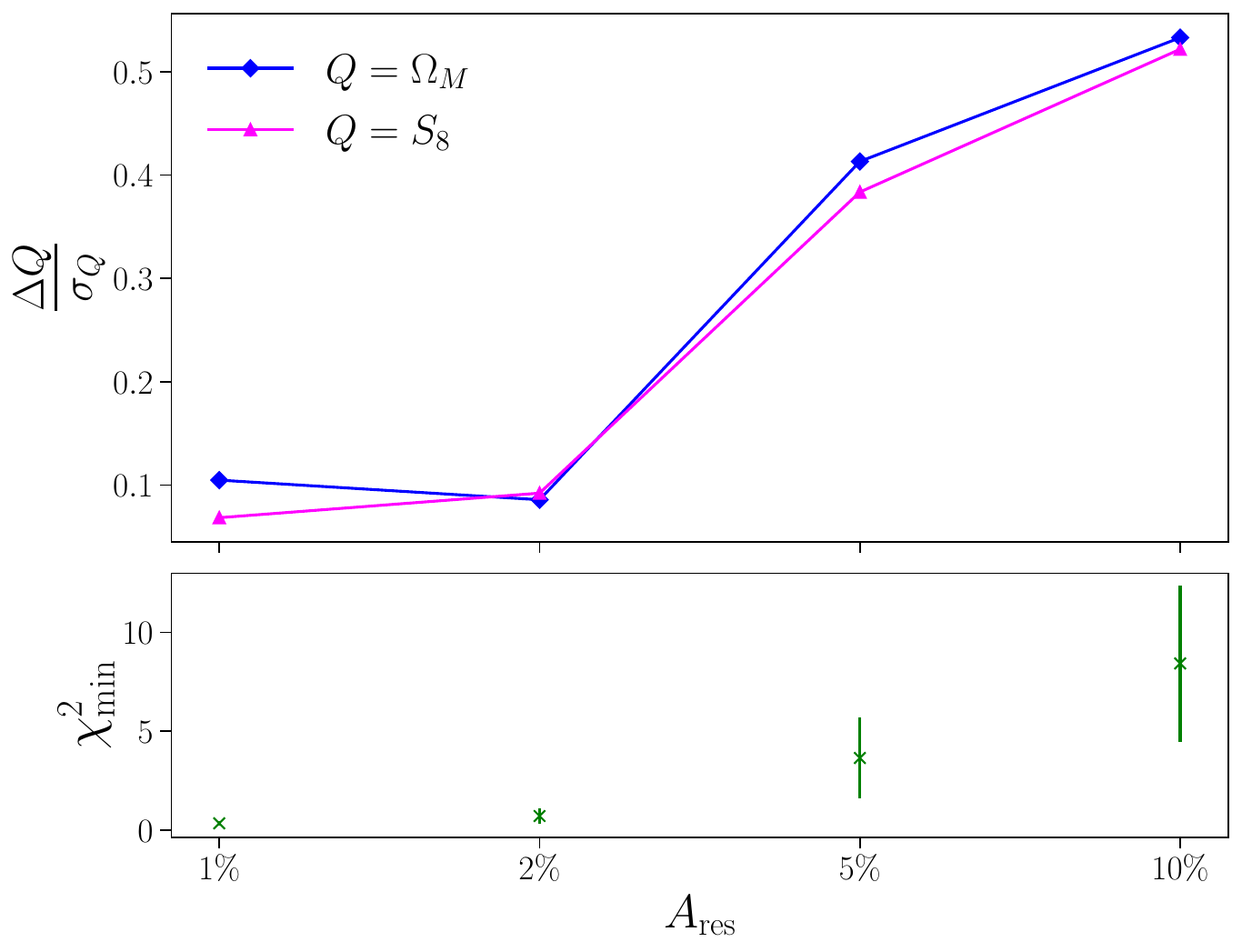}
        \caption{Same as in Fig. \ref{fig:gaussian-NLA} but assuming the TATT model in the analysis. Note that the lower bias on $\Omega_M$ at $\resamp=2\%$ with respect to $\resamp=1\%$ is purely statistical due to the limited number of GP realisations used here.}
        \label{fig:gaussian-TATT}
    \end{figure}

    \begin{figure}
      \includegraphics[width = \columnwidth]{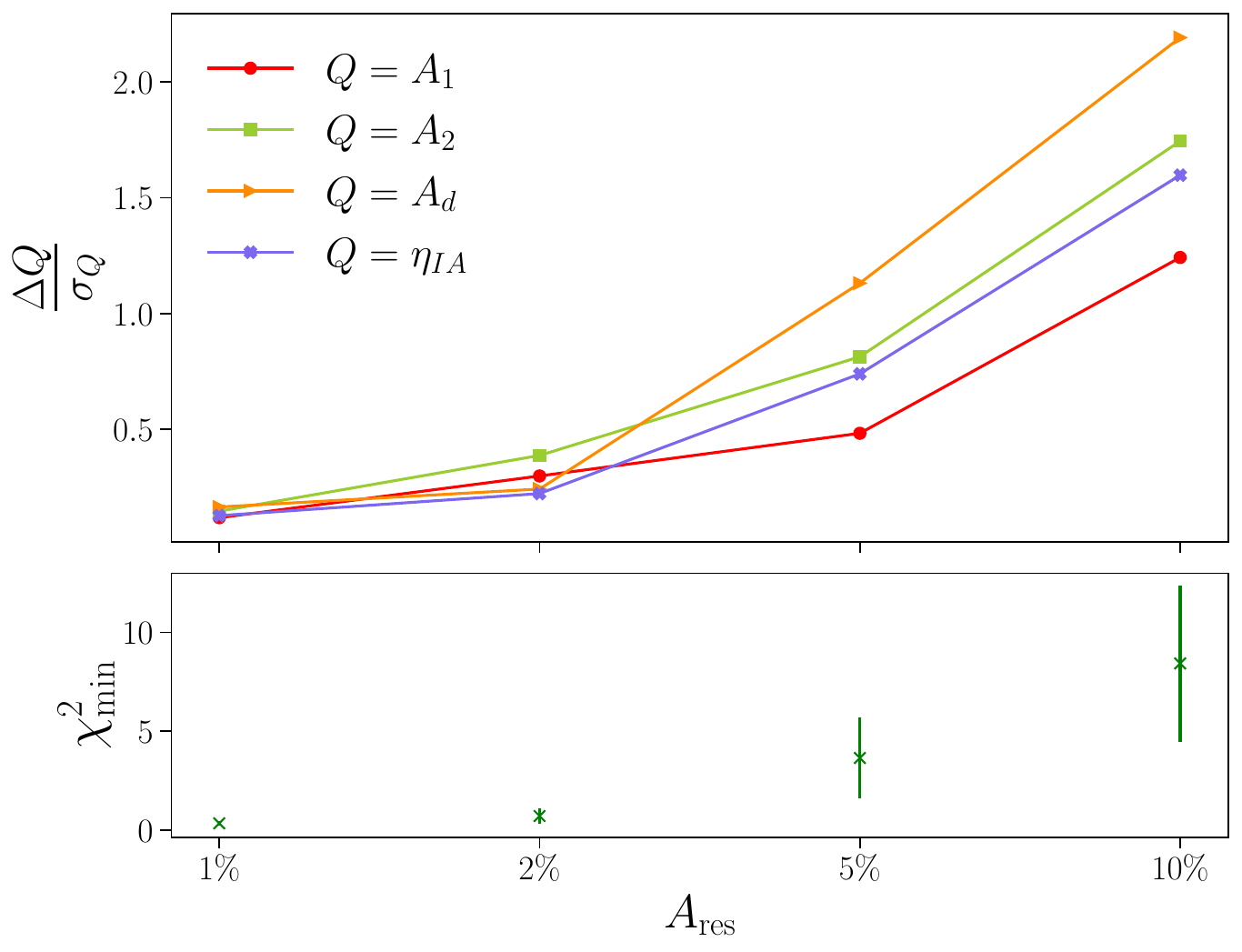}
      \caption{Average relative bias of the TATT parameters for the same analysis as in Fig. \ref{fig:gaussian-TATT}.}       \label{fig:gaussian-TATTparams}
    \end{figure}

    Figure \ref{fig:gaussian-TATT} shows the result of same exercise using TATT to analyse the perturbed data vector. Similarly to the NLA case, the relative bias increases with the amplitude of the perturbation but remains below $\sim0.5\sigma$ for a up to a $\resamp=10\%$ residual amplitude. Figure \ref{fig:gaussian-TATTparams} shows that, in turn, the IA TATT parameters are significantly biased (between $1$ and $\sim2\sigma$ for $\resamp=0.1$). This is interesting, since it shows that the more flexible TATT model tries to absorb the IA residual by changing its free parameters, but this does not translate into a significant change in the bias of the cosmological parameters. This is a manifestation of the markedly different redshift dependence of IAs compared to weak lensing. While the lensing signal is cumulative and thus correlates a given redshift bin with all other bins at some level, IAs are local in redshift, and only correlate with the IA signal in adjacent bins, or with the lensing signal from bins at higher redshifts. It is thus harder (although not impossible) to find variations of the cosmological parameters that are able to accommodate IA modelling residuals, whereas this is not the case for parameters that only modify the IA contribution.

    It is not surprising that, despite this additional freedom, TATT is not able to significantly reduce the parameter bias in the case of GP-like IA residuals. As we discussed in the previous section, the GP model is unphysical in that it generates perturbations at all scales, and with an arbitrary scale dependence, whereas deviations from NLA, certainly those that TATT would be able to absorb, are expected to occur on smaller scales, and have a relatively smooth scale dependence. This simpler behaviour could also be easier to mimic by a variation in cosmological parameters, and hence we must also study the impact of IA residuals with more physically-motivated properties.
    
  \subsection{Simulation-based perturbations} \label{ssec:res.sim}
    We now explore the simulation-based model for the IA residuals introduced in Section \ref{sssec:meth.pert.sim}, and parametrised as in Eq. \ref{eq:fitfunc}. We vary parameters of the fitting function, $k_p$, $\Delta k$ and $\resamp$, and estimate the relative bias on $\Om$ and $S_{8}$. In the following subsections, we present the results found under different assumptions regarding the IA model used in the analysis.

    \begin{figure}
        \centering
        \includegraphics[width = \columnwidth]{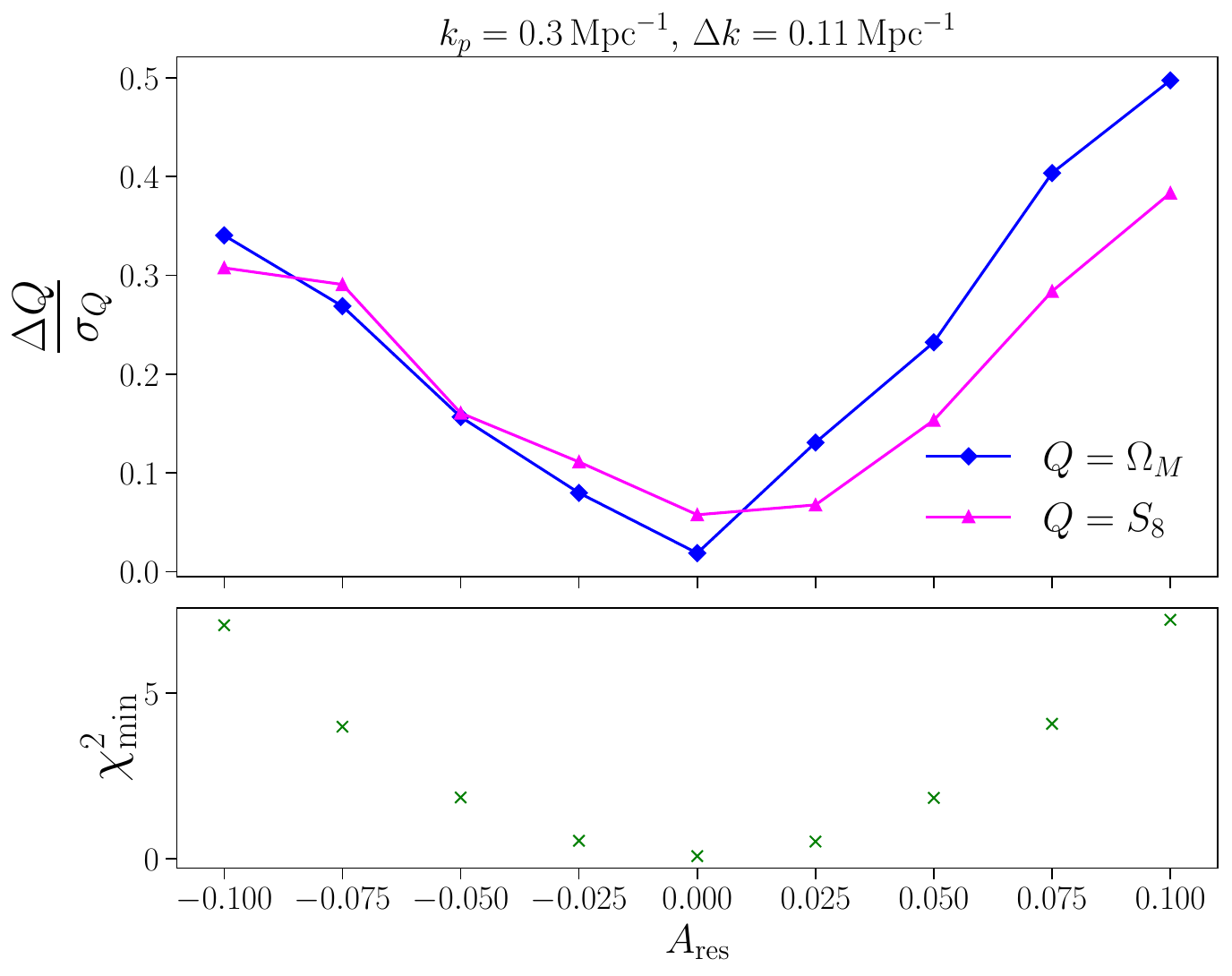}
        \includegraphics[width = \columnwidth]{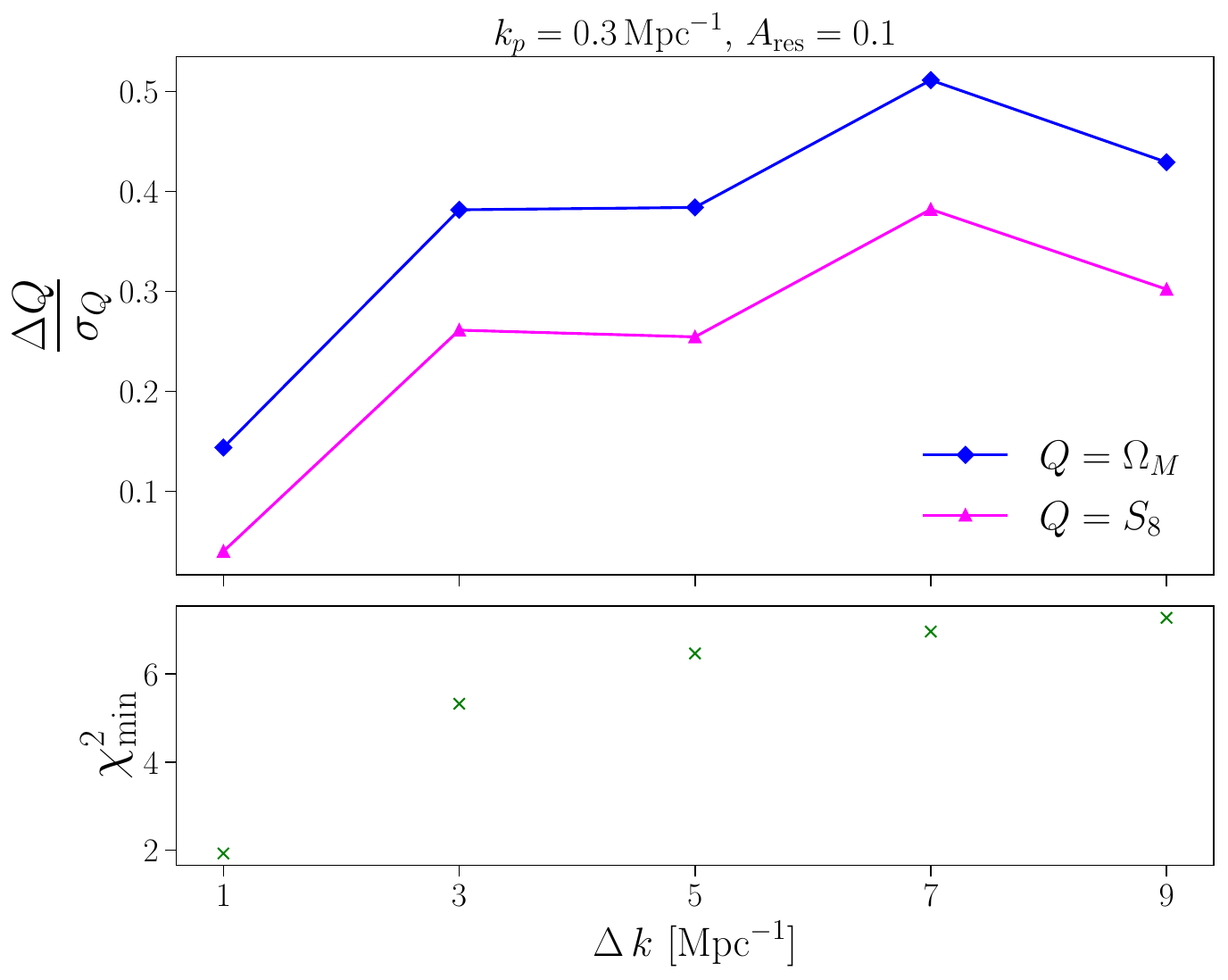}
        \includegraphics[width = \columnwidth ]{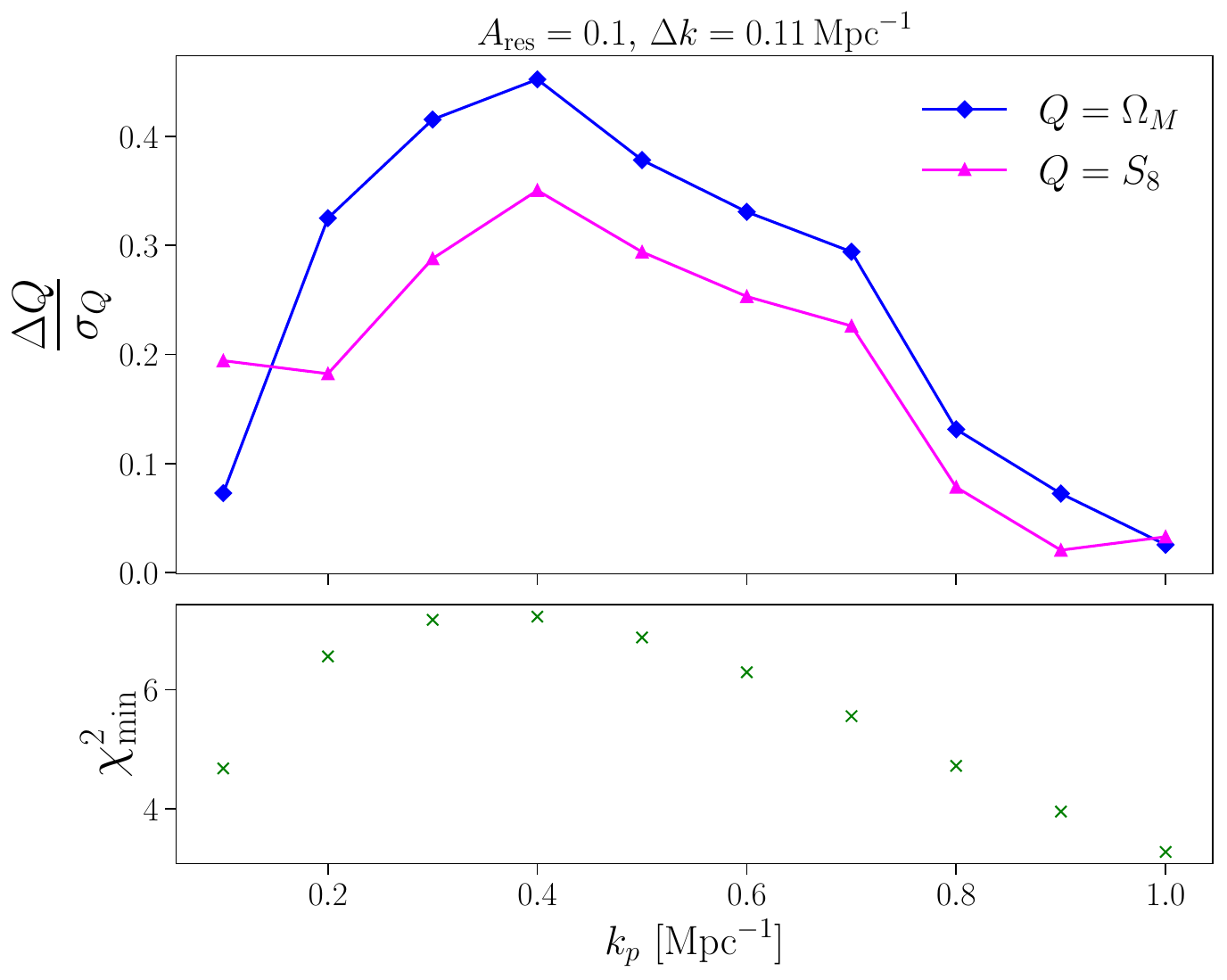}
        \caption{Relative bias in $\Om$ and $S_8$ from data with IA residuals generated using the simulation-based model of Eq. \ref{eq:fitfunc}. In all panels, the fiducial values of the pivot scale, gradient scale, and residual amplitude are $k_p=0.3\,{\rm Mpc}^{-1}$, $\Delta k=0.11\,{\rm Mpc}^{-1}$, and $\resamp=0.1$, respectively, and we show the results of varying one of these parameters at a time. The dependence on $\resamp$, $\Delta k$, and $k_p$ is shown in the top, middle, and bottom panes, respectively. The bottom sub-panels shows the minimum $\chi^2$ achieved in each case, as a goodness-of-fit metric.}
        \label{fig:phys_varied}
    \end{figure}
    
    \subsubsection{Assuming NLA}
      Analysing data with IA residuals generated according to Eq. \ref{eq:fitfunc}, with the fiducial parameters found from the measurements of \cite{Shi2021} $\{\resamp=0.1,\,k_p=1.2\,{\rm Mpc}^{-1},\,\Delta k=0.11\,{\rm Mpc}^{-1}\}$, causes only a small bias on cosmological parameters: 0.064$\sigma$ for $\Om$ and 0.006$\sigma$ for $S_8$. This is because, as found by \cite{Shi2021}, the IA residuals with respect to the NLA model only kick in on relatively small scales ($k\gtrsim1\,{\rm Mpc}^{-1}$), to which we are less sensitive given the scale cuts assumed in this analysis, and the larger measurement uncertainties on those scales due to the contribution from shape noise (see Fig. \ref{fig:a_ia_fit}).

      For this reason, in what follows, we will proceed by changing the fiducial value of $k_p$ to $k_p=0.3\,{\rm Mpc}^{-1}$, such that the deviations with respect to a pure NLA model kick in on relatively large scales (while preserving a scale dependence based on simulated data). With this new pivot scale, and keeping $\resamp$ and $\Delta k$ to their fiducial values, the parameter biases are significantly larger: $0.4\sigma$ and $0.3\sigma$ for $\Om$ and $S_8$, respectively. This is of the same order as the relative biases obtained for the Gaussian process residuals with 10\% amplitude. With this in mind, let us now explore the impact of varying the values of $\{\resamp,\,k_p,\,\Delta k\}$ in order to study the sensitivity of this result on the specific scale dependence of the IA residuals.

      The results are shown in Fig. \ref{fig:phys_varied}, which displays the relative biases on $\Om$ and $S_8$ as a function of $\resamp$,  $\Delta k$, and $k_p$ (top, center, and bottom panels respectively). As shown in the top panel, for residual amplitudes in the range $|\resamp|\leq0.1$, the parameter biases grow in direct proportion to $|\resamp|$, remaining below $\Delta Q/\sigma_Q\lesssim0.5$. The best-fit $\chi^2$ (shown in the bottom subpanels of each figure) is, in all cases, small enough compared to $N_{\rm dof}$, such that the IA residuals would not be detected through a poor fit of the model to the data.
      \begin{figure}
          \centering
          \includegraphics[width = \columnwidth]{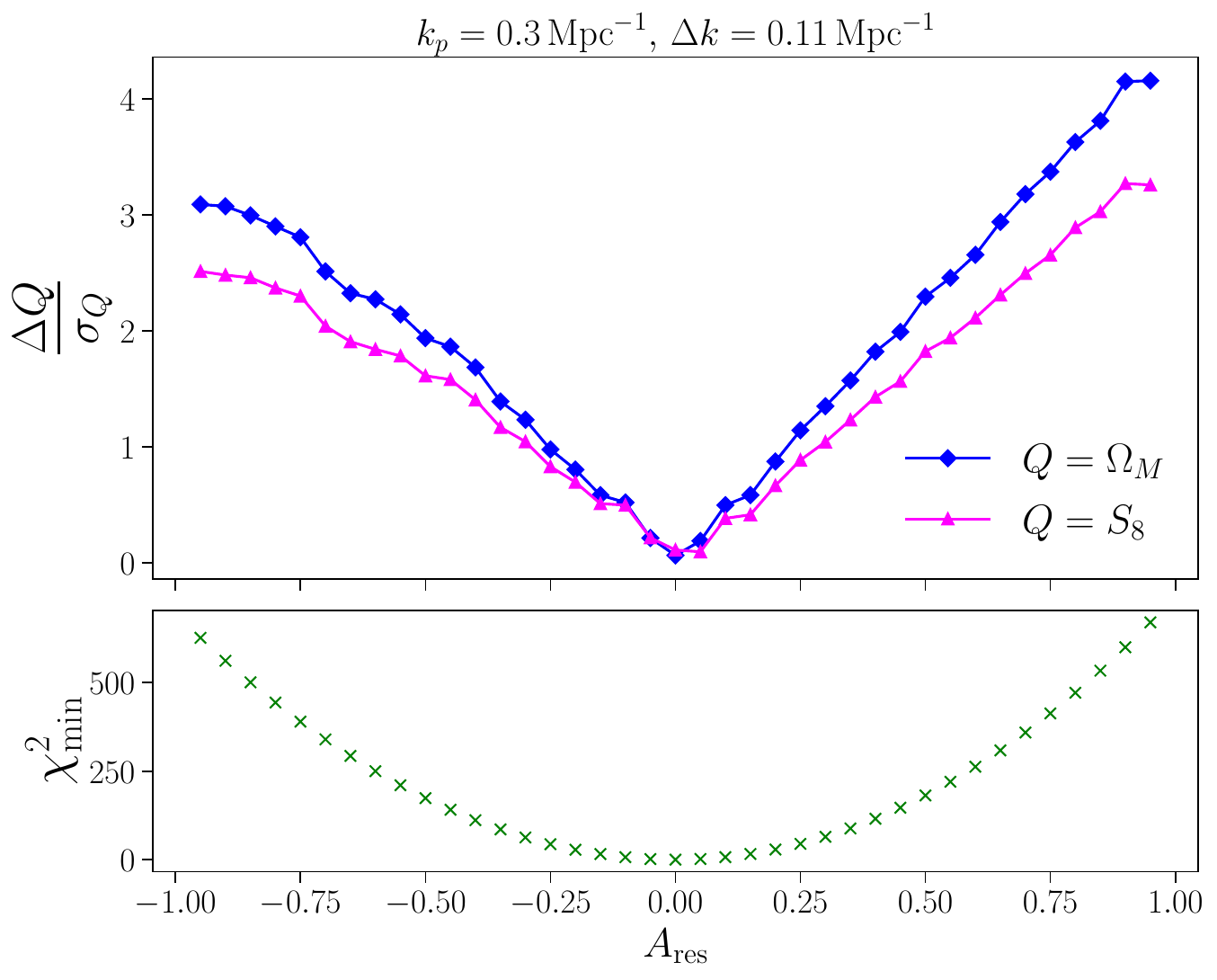}
          \caption{Relative bias on cosmological parameters for larger amplitudes of the simulation-based IA residuals than those explored in Fig. \ref{fig:phys_varied}, assuming an NLA model in the analysis. The bottom panel shows the minimum $\chi^2$ achieved in each case, as a goodness-of-fit metric.}
          \label{fig:sim-break}
      \end{figure}

      The middle panel shows the relative biases found as a function of $\Delta k$ in the range $0.1\,{\rm Mpc}^{-1}<\Delta k<1\,{\rm Mpc}^{-1}$. $\Delta k$ describes the gradient of the transition between the unperturbed NLA IA power spectrum and the perturbed one. The bias on the cosmological parameters decreases with $\Delta k$, and the $\chi^2$ values plateau towards $\chi^2\sim7$. This is to be expected, as increasing $\Delta k$ smooths the transition between the unperturbed and perturbed regimes of the IA power spectrum, widening the range of scales over which the model can fit the data. This behaviour (growing towards low values of $\Delta k$) is reproduced by both the relative bias and the best-fit $\chi^2$ bias. We repeated the experiment varying $k_p$ within the range $[0.1,1.0]$, with the results shown in the bottom panel of Fig. \ref{fig:phys_varied}. The maximum bias for both $\Om$ and $S_8$ ($\sim0.5\sigma$ and $\sim0.35\sigma$ respectively) is attained at $k_p = 0.4 \,\,{\rm Mpc}^{-1}$, where the best-fit $\chi^2$ is also the largest. This confirms that the fiducial value of $k_p=0.3\,{\rm Mpc}^{-1}$ used here is indeed a conservative choice. The behaviour otherwise is easy to understand: at the smallest $k_p$ scales, we essentially expose most of the power spectrum to the IA residual which, in this case, effectively amounts to an overall multiplicative factor, easily fitted by $A_1$. On the other hand, as we raise $k_p$, we eventually push the IA perturbation outside of the range of scales to which our data are sensitive.

      So far the conclusion seems to be that IA modelling residuals at the $\sim10\%$ level cause only moderate shifts in the final cosmological parameters ($\lesssim0.5\sigma$). In order to quantify under what conditions this conclusion breaks down, we have repeated the same exercise, varying the residual amplitude $\resamp$ within significantly wider bounds ($\resamp\in[-1,1]$). The results are shown in Fig. \ref{fig:sim-break}. We find that a $25 \%$ residual from the assumed IA power spectrum produces a bias of $1\sigma$ for both cosmological parameters. At that point, however, the residual $\chi^2$ value is $\chi^2=44$, corresponding to probability-to-exceed value ${\rm PTE}=0.055$. Although this is a relatively low value, it would likely not be sufficient to pinpoint the IA model as the main source of residual systematics. For $\resamp=0.42$ (i.e. $\sim40\%$ residuals, corresponding to $\sim1.5\sigma$ parameter biases), the PTE drops sharply to $\sim0.001$, which would be a definitive indication of a modelling failure. It is worth noting, however, this conclusion may depend on the specific analysis choices made here (e.g. scale cuts, or the assumption of a purely Gaussian covariance matrix).

    \subsubsection{Assuming higher-order models}
      \begin{figure}
        \centering
        \includegraphics[width = \columnwidth]{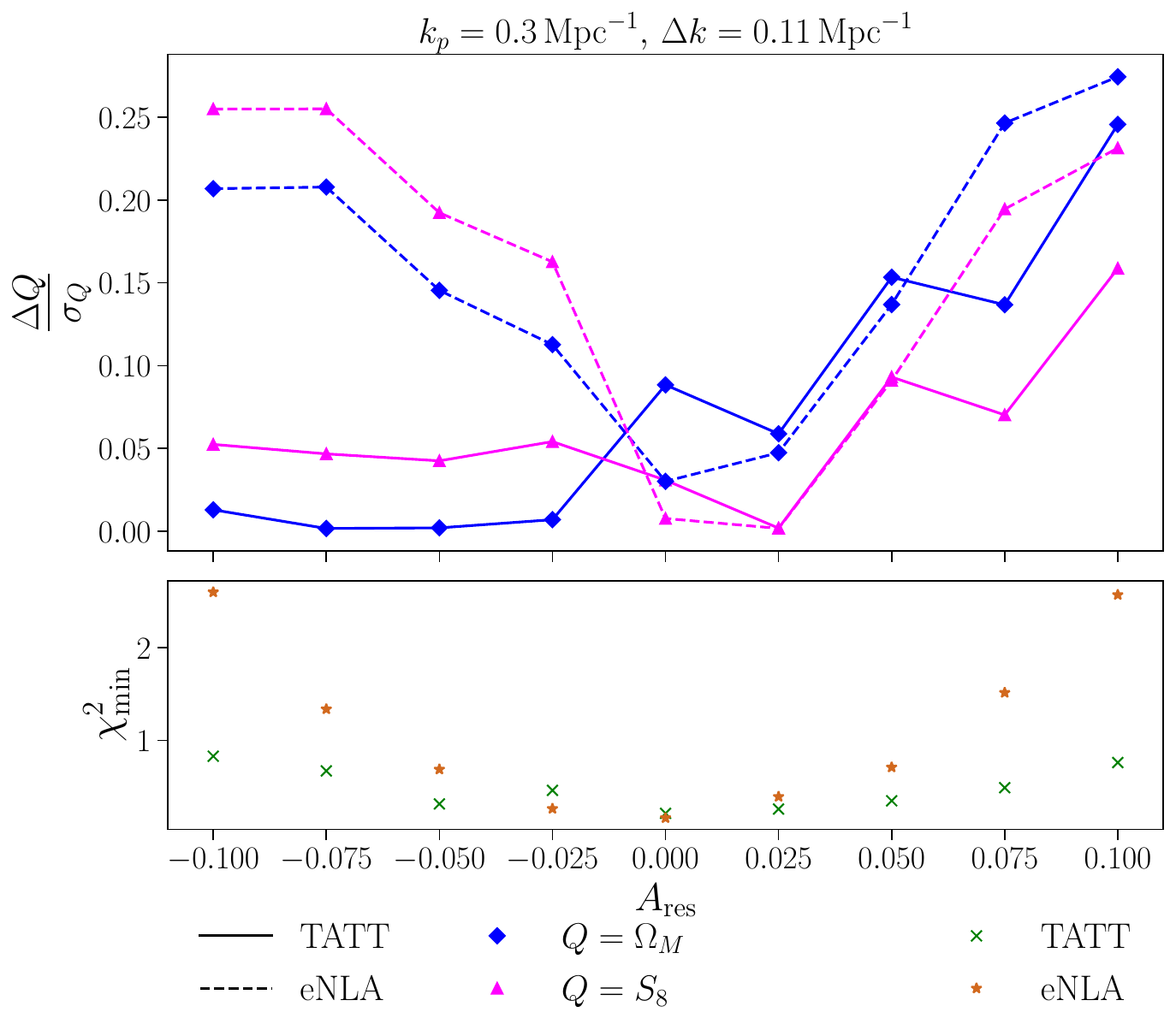}
        \caption{Relative bias on $\Om$ and $S_8$ from data with simulation-based IA residuals analysed with the TATT (solid lines) and eNLA (dashed lines) models, as a function of the residual amplitude $\resamp$. The bottom panel shows the minimum $\chi^2$ achieved in each case, as a goodness-of-fit metric.}
        \label{fig:simNLA-TATT}
      \end{figure}
      \begin{figure}
        \centering
        \includegraphics[width = \columnwidth]{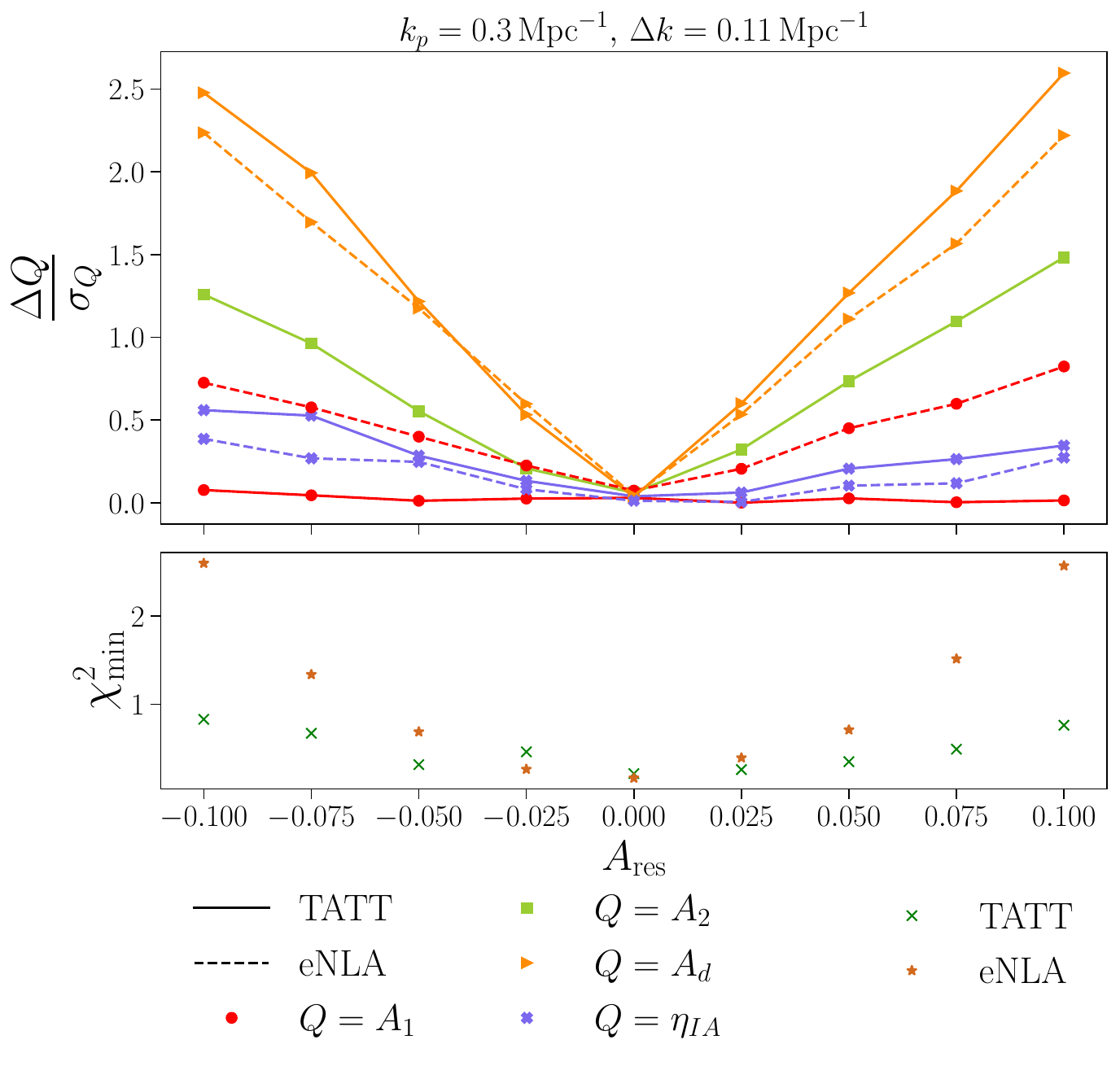}
        \caption{As Fig. \ref{fig:simNLA-TATT} for the relative bias on the TATT/eNLA parameters.}
        \label{fig:simNLA-TATTparams}
      \end{figure}
      We repeated the analysis from the previous section but now assuming the TATT intrinsic alignment model, with 2 additional free parameters ($A_2$ and $A_{1\delta}$), to analyse the data. The solid lines in Fig. \ref{fig:simNLA-TATT} show the resulting cosmological parameter biases as a function of the residual amplitude $\resamp$. In this case, TATT incurs parameter biases of up to $0.25\sigma$ and $0.15\sigma$ for $\Om$ and $S_8$, respectively. These are significantly smaller than the NLA case ($0.4\sigma$ and $0.3\sigma$), due to the larger flexibility of TATT to absorb the IA residuals. Although this is in contrast with the results found in the case of Gaussian process residuals, the result is not surprising: while the GP-induced residuals induce perturbations on all scales, the simulation-based model has a much simpler scale dependence that should be easier to mimic by perturbative expansions in which higher-order terms are more relevant on smaller scales. The solid lines in Fig. \ref{fig:simNLA-TATTparams} show the relative biases in the TATT IA parameters with respect to their fiducial, unperturbed values ($A_1=1$, $A_2=A_{1\delta}=0$, $\eta_1=0$) as a function of $\resamp$. As the plot shows, the density-weighting amplitude $A_{1\delta}$ incurs the largest bias, indicating that it absorbs most of the IA residuals.

      The results above, together with the lack of detection of a non-zero $A_2$ in both simulations and real data, motivates an exploration of the simpler eNLA model, in which the quadratic tidal field term is set to $A_2=0$, while $A_{1\delta}$ is allowed to vary (effectively representing the weighting of the IA signal by the local galaxy overdensity). The resulting relative biases on cosmological parameters, as a function of $\resamp$ are shown as dashed lines in Fig. \ref{fig:simNLA-TATT}. Interestingly, we observe that, within the range of $\resamp$ explored here, the maximum relative bias is only marginally larger than that achieved with the more general TATT parametrisation. At the same time, fixing $A_2=0$ seemingly restores the symmetry of this curve around $\resamp=0$. We interpret this as due to a significant degeneracy between $A_1$ and $A_2$ for the data vector considered here, which allows $A_2$ to absorb some of the large-scale IA power caused by the residual in an asymmetric way. This is more evident after considering the relative bias to the IA parameters incurred by the eNLA model, shown as dashed lines in Fig. \ref{fig:simNLA-TATTparams}. We can see that, after fixing $A_2=0$, $A_{1\delta}$ is still as biased as it was in the TATT case, whereas $A_1$ now becomes significantly more biased, having absorbed most of the residual that was previously captured by $A_2$. The full set of 2D parameter contours obtained in this particular exercise are shown in Fig. \ref{fig:triangle_full} in Appendix \ref{app:triangle}.

      In summary, we find that perturbative IA expansions are able to capture more of the parameter bias caused by residual fluctuations in the IA signal with a physically motivated scale dependence, and that the performance of simpler perturbative models, such as the eNLA parametrisation, is comparable to more complex parametrisations given the experimental sensitivity assumed here.
    
  \subsection{Other physically-motivated parametrisations}
    To further test the robustness of the results obtained in the previous sections, here we analyse simulated data vectors with residual IA signal generated with alternative physically-motivated prescriptions.

    We first explore the halo model parametrisation described in \ref{sssec:meth.pert.hm}. With a residual amplitude $\resamp=0.1$, and assuming the NLA model in the likelihood, we find a relative bias of $0.14\sigma$ on $\Om$, and $0.08\sigma$ on $S_8$. When analysed under the TATT parametrisation, these biases are further reduced to $0.09\sigma$ for $\Om$ and $0.015\sigma$ for $S_8$. This is comparable to the parameter biases observed for the simulation-based parametrisation with fiducial parameters discussed in the previous section.

    Secondly, we generate TATT-like residuals as described in Section \ref{sssec:meth.pert.tatt}, for a 10\% residual amplitude. Analysed under the NLA model, this perturbed data vector recovers a bias of $0.02\sigma$ for $\Om$ and $0.12\sigma$ for $S_8$. This is in reasonable agreement with the results found for the halo model and the simulation-based residuals with a similar amplitude.

    Finally, since so far we have only explored residuals with respect to a baseline NLA-like signal with $A_1=1$, we also study the possibility of adding a physically-motivated residual on a baseline IA power spectrum predicted by a TATT model with values of $A_2$ and $A_{1\delta}$ that are significantly different from zero. In particular, we construct a baseline TATT model fixing $A_1=1$, and choosing values for $A_2$ and $A_{1\delta}$ such that the resulting power spectrum deviates by up to $50\%$ from the NLA component (i.e. the model with $A_2=A_{1\delta}=0$) over the range of scales explored here. We then take the resulting power spectrum, and add a simulation-based residual using Eqs. \ref{eq:fitfunc} and \ref{eq:simii}, with a residual amplitude $\resamp=0.1$, and a conservative pivot scale $k_p=0.3\,{\rm Mpc}^{-1}$. The resulting data vector is then fitted with the TATT model. The aim of this exercise is to verify that a baseline IA model that departs significantly from NLA would not change the results obtained in Section \ref{ssec:res.sim} (e.g. due to a non-trivial interplay between the scale dependence induced by NLA and that of the simulation-based residual). The parameter biases recovered in this case are $0.1\sigma$ and $0.18\sigma$ for $\Om$ and $S_8$, respectively, in reasonable agreement with the results displayed in Fig. \ref{fig:phys_varied} when using NLA as the baseline, unperturbed, IA signal.

  \subsection{The impact of the overall IA amplitude}\label{ssec:res.amp}
    \begin{figure}
        \centering
        \includegraphics[width = \columnwidth]{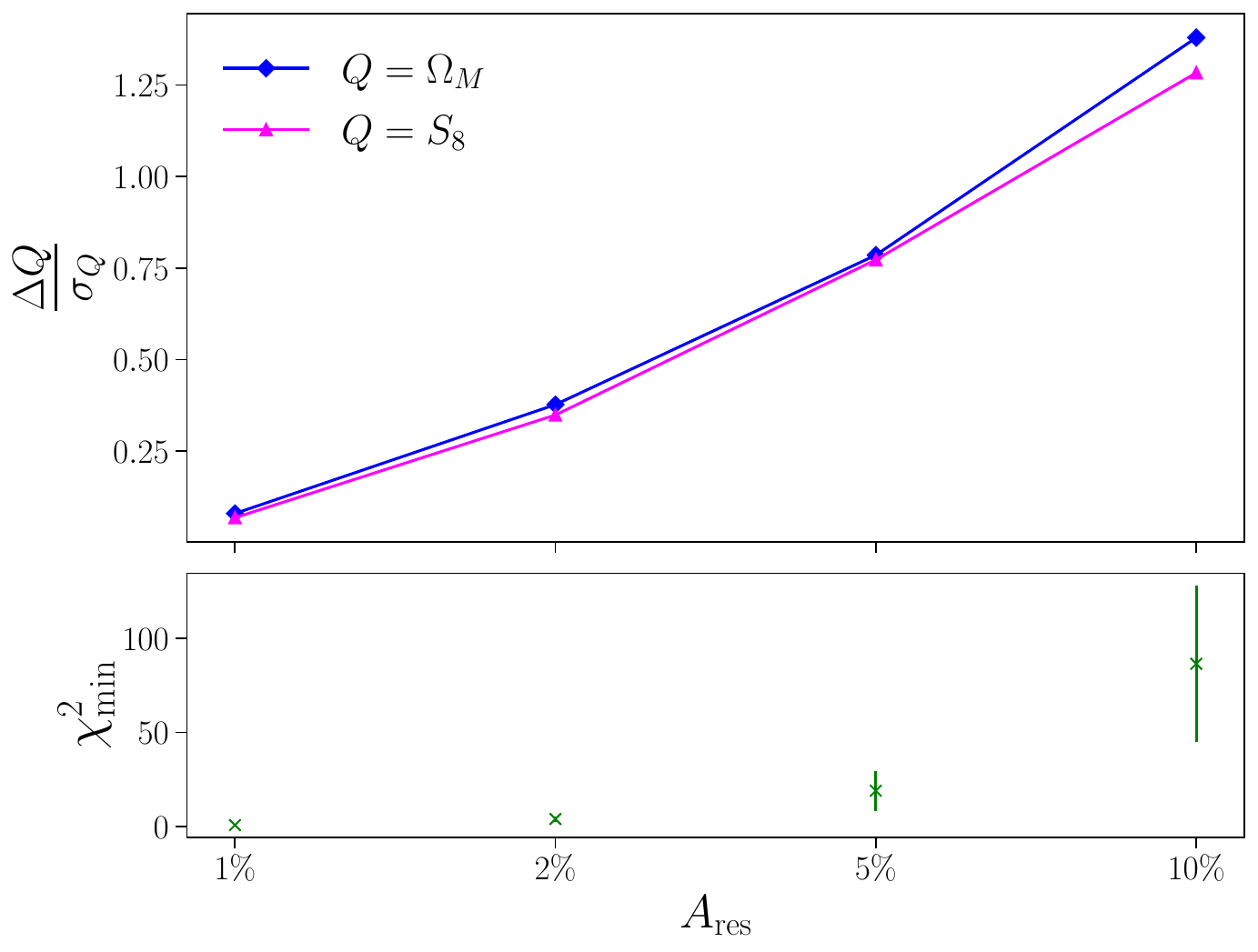}
        \caption{Average bias of $\Om$ and $S_8$ from Gaussian process residuals over a baseline IA model with amplitude $A_\mathrm{IA}=2$, when assuming the NLA model in the analysis, as a function of the residual amplitude $\resamp$. The bottom panel shows the minimum $\chi^2$ achieved in each case, as a goodness-of-fit metric.}
        \label{fig:gaussian-A2}
    \end{figure}
    \begin{figure}
        \centering
        \includegraphics[width = \columnwidth]{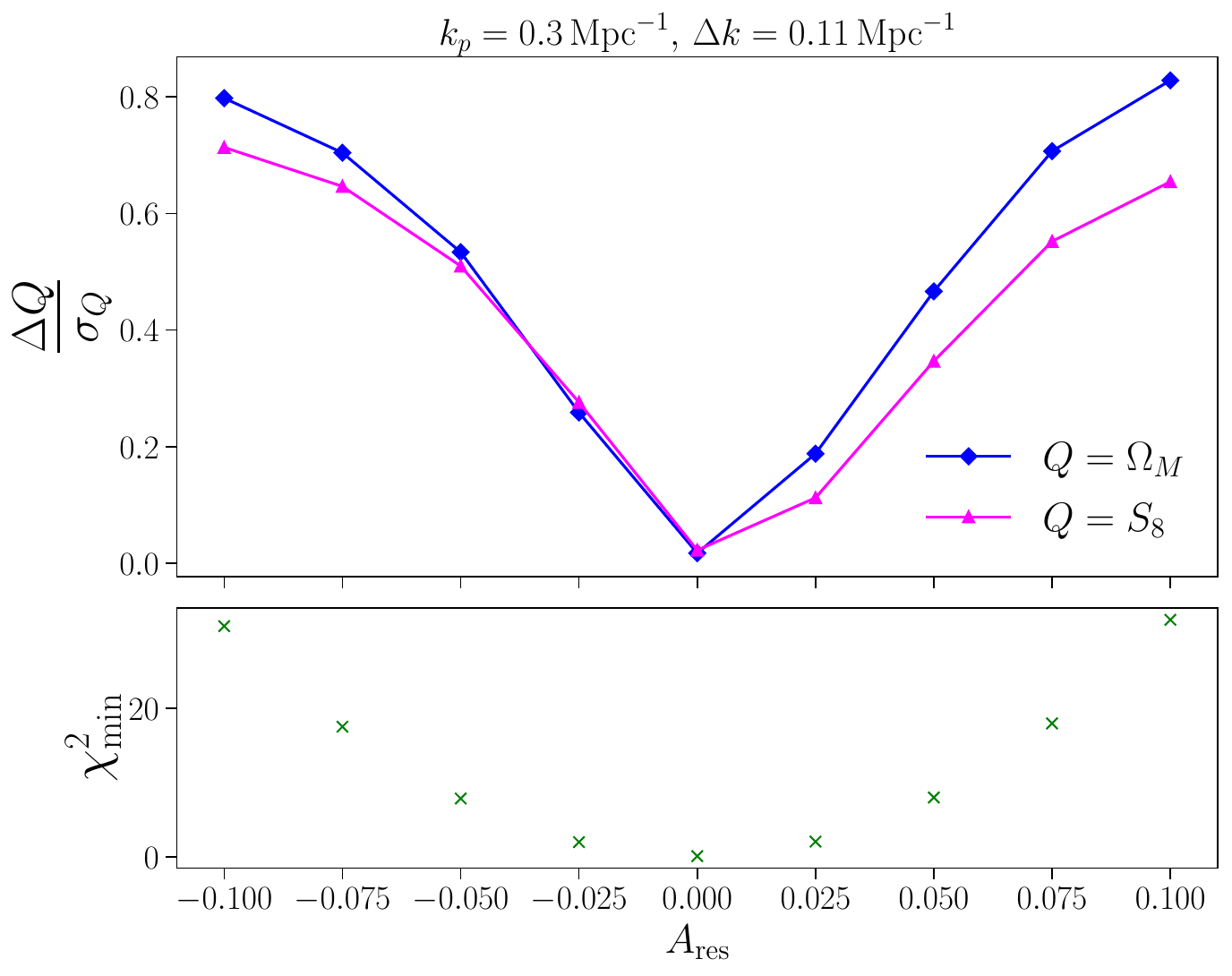}
        \caption{As Fig. \ref{fig:gaussian-A2} for data with simulation-based IA residuals.}
        \label{fig:phys-A2}
    \end{figure}
    The qualitative result found so far is that modelling residuals with respect to the baseline IA model, with amplitudes of up to $\sim10\%$, cause biases on cosmological parameters that are, at most, about $0.5\sigma$. As argued in Section \ref{sssec:meth.pert.hm}, this is directly related to the ratio between the amplitude of the IA contribution to the total shear signal, and the size of statistical uncertainties expected from Stage-IV surveys. Thus, this result depends strongly on the assumed baseline IA amplitude.
    
    We have so far adopted a baseline IA signal described by the NLA model with $A_1=1$, which sets the overall amplitude of the IA contribution. This choice corresponds roughly to the highest value of $A_1$ measured by current weak lensing surveys (specifically, from the bandpower analysis of KiDS-1000 data - see \cite{Asgari2021}), and should therefore be a conservative estimate of the typical IA amplitude expected in future surveys. This, however, depends strongly on the specifics of the source sample under study. For instance, red galaxies are known to exhibit significantly stronger tidal alignments, with $A_1\gtrsim5$ \citep{Singh2015, Fortuna2021}, and hence the red fraction of a given source sample plays a major role in determining the amplitude of the IA contribution.

    Following the rationale above, it would be reasonable to expect that, for a given experiment, with fixed statistical uncertainties, the parameter biases induced by IA residuals with a given relative amplitude would scale in direct proportion to the overall amplitude of the baseline IA contribution (i.e. $\propto A_1$). To test this, we repeat the analysis of Gaussian process and simulation-based residuals adopting a baseline NLA model with $A_1=2$. The resulting parameter biases as a function of the residual amplitude $\resamp$ are shown in Fig. \ref{fig:gaussian-A2} for Gaussian processes, and Fig. \ref{fig:phys-A2} for the simulation-based model. Visual comparison with Figs. \ref{fig:gaussian-NLA} and \ref{fig:phys_varied} confirms our rough expectation: parameter biases double with respect to those found in previous sections with an overall IA signal amplitude $A_1=1$. This serves as a warning that the results presented here depend directly on the IA amplitude assumed, while at the same time providing a simple rule of thumb to extrapolate our results for samples with different IA amplitudes.

\section{Conclusions} \label{sec: conclusions}
  Together with the impact of baryonic effects in the non-linear matter power spectrum, intrinsic alignments are a key source of astrophysical modelling uncertainty for current and future weak-lensing surveys. Since a detailed model of intrinsic alignments derived from first principles is far beyond our current understanding of the physics of this phenomenon, it is key to quantify the level of accuracy we must achieve for their modelling so as not to impact the robustness of cosmological constraints from weak lensing data. This has been the goal of this paper: not to discriminate between the ability of specific models to describe IAs, but to quantify the level to which any such model must be able describe the IA signal.

  To do this, we have generated synthetic two-point cosmic shear data contaminated by IAs featuring residual contributions not described by any specific model. To ensure that our results are not dependent on the prescription used to generate these residuals, we have explored a variety of methodologies, ranging from simulation-based (and hence physically motivated) approaches, to completely agnostic methods based on Gaussian processes. The general result derived from this study is the following: for an LSST-like survey, representative of future Stage-IV experiments, and assuming a baseline IA amplitude consistent with the highest so far detected in cosmic shear samples, the IA contribution to shear two-point correlation functions must be modelled at the $\sim10\%$ accuracy level to avoid biasing final constraints on cosmological parameters by more than $\sim 0.5\sigma$. This parameter bias may be significantly smaller, depending on the scale dependence of these residuals, and on the model used to analyse the data.

  Two important consequences of this result must be emphasized. First, a $\sim10\%$ accuracy requirement is relatively modest (e.g. compared to the usual percent-level modelling requirement of most core cosmological observables), and could be achievable by sufficiently flexible modelling frameworks already in existance, such as perturbative expansions \cite{Blazek2019}, halo-based models \cite{Fortuna2021}, or hybrid parametrisations \cite{Maion2023}. This in principle bodes well for future cosmological constraints from weak lensing. Secondly, however, this level of accuracy must be achieved over the whole range of scales to which cosmic shear data is sensitive, which include highly non-linear scales ($k\sim O(1\,{\rm Mpc}^{-1})$), on which perturbative approaches may fail.

  We have shown that our results are robust to the detailed properties of the modelling residuals assumed, as well as to the choice of imperfect IA model used to analyse the synthetic data. Nevertheless, a number of caveats must be noted that could affect these results, and which could be addressed in follow-up studies. First, as we noted in Section \ref{ssec:res.amp}, the accuracy requirement depends crucially on the overall size of the IA signal, and could become significantly tighter for source samples displaying larger alignment amplitudes. This can be remedied through a judicious design of this sample (e.g. avoiding early-type galaxies, which have been shown to display stronger alignments). The requirement found also depends on the range of scales used in the analysis, and thus could become significantly tighter for studies targeting the smallest angular scales scales (although other important astrophysical modelling uncertainties will also affect such analyses). Our analysis has focused only on cosmic shear correlations, and ignored the combination with tomographic galaxy clustering (the so-called ``3$\times$2-point'' analysis). The smaller statistical uncertainties achieved in this extended analysis could tighten the $\sim10\%$ modelling requirement found here, particularly in respect to the IA contamination to the galaxy-galaxy lensing signal. Finally, we have only considered a limited range of cosmological and nuisance parameters in our analysis ($\Om$, $S_8$, and the IA model parameters). The reason for this was twofold: first, a limited model would avoid certain parameter degeneracies and achieve tighter constraints on cosmological parameters, thus leading to a more conservative IA modelling requirement. Secondly, a reduced parameter space allowed us to run the large number of MCMC chains required for this analysis in a reasonable timescale. Our analysis has thus ignored potential degeneracies of the cosmological parameters with the IA parameters caused by other nuisance parameters, such as photometric redshift uncertainties, as well as the potential for other cosmological parameters (e.g. the Dark Energy equation of state) to be more affected by IA residuals. Since our results seem to be largely driven by the size of the overall IA signal in relation to the statistical uncertainties in the data, we do not expect this simplification to significantly affect our results. However, we do note in particular that a there is growing evidence for degeneracy between parameters of IA and of photometric redshift uncertainty models \citep{wright2020kids, stolzner2021self, fischbacher2023redshift}, so future work in this vein which considered these systematic effects jointly may be of interest.

  In any case, ultimately this study should be followed-up by an in-depth analysis that employs the modelling requirements found here to quantify the suitability of different IA parametrisations for future weak lensing surveys. Similar studies have been carried out in the context of other astrophysical systematics, such as galaxy bias \citep{Nicola2023}, cosmic magnification \citep{Lorenz2018,Duncan2022}, and higher-order corrections to the lensing power spectrum \citep{Deshpande2023}. Although challenging, the requirements derived here give hope that the impact of IAs in cosmological weak-lensing analyses can be kept under control in the Stage-IV era.

\section*{Acknowledgements}
  We thank Raul Angulo, Jonathan Blazek, and Elisa Chisari for useful comments and discussions. We also thank Ashling Gordon and Jonathan Patterson for administrative and technical support that was essential for this project. AP is partially supported by the National Astronomical Institute of Thailand (NARIT) and St Peter's College, Oxford. NP and DA acknowledge support from the Beecroft Trust, while JHD acknowledges support from an STFC Ernest Rutherford Fellowship (project reference ST/S004858/1). We made extensive use of computational resources at the University of Oxford Department of Physics, funded by the John Fell Oxford University Press Research Fund.

\bibliography{main}
\appendix

\section{Full parameter constraints}\label{app:triangle}
\begin{figure}
    \centering
    \includegraphics[width = 0.9\linewidth]{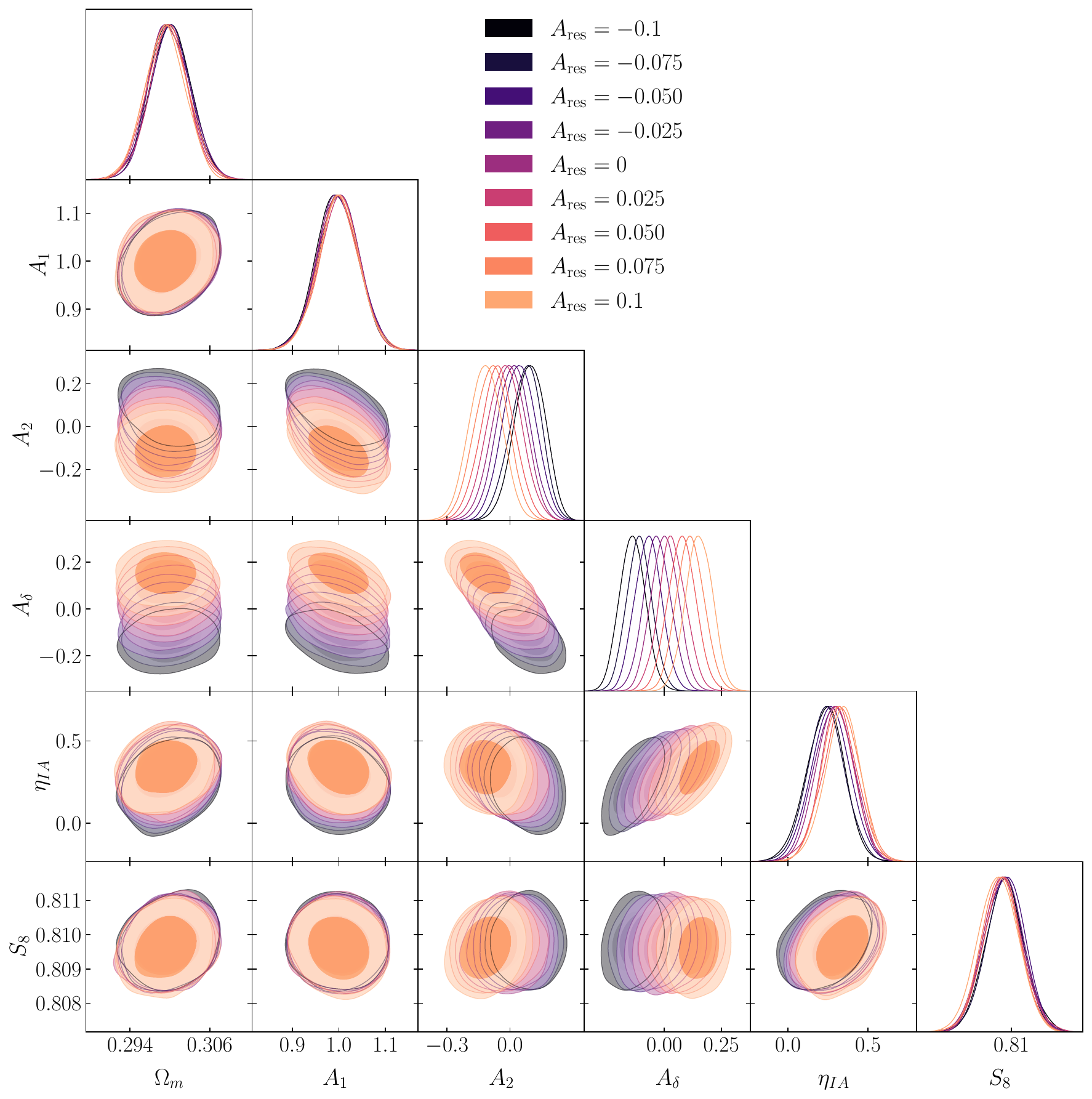}
    \caption{Posteriors of TATT parameters on data generated with NLA and simulation-based perturbations ranging from $\resamp \in [-0.1, 0.1]$}
    \label{fig:triangle_full}
\end{figure}

Fig. \ref{fig:triangle_full} shows the complete set of 2D parameter contours for the cosmological ($\Om$, $S_8$) and IA parameters ($A_1$, $A_2$, $A_{1\delta}$, $\eta_1$) for a synthetic data vector with simulation-based IA residuals ($k_p=0.3\,{\rm Mpc}^{-1}$, $\Delta k=0.11\,{\rm Mpc}^{-1}$) with different residual amplitudes, ranging from $\resamp=-0.1$ to $\resamp=0.1$, and analysed assuming the TATT model. As a reminder, the maximum relative bias on cosmological parameters, displayed here, corresponds to $0.25\sigma$ for $\Om$ and $0.15\sigma$ for $S_8$.
\end{document}